\documentclass[preprint]{ptephy}

\preprintnumber{KYUSHU-HET-183}

\usepackage{amssymb}
\usepackage{amsthm}
\usepackage{amsmath}
\usepackage{booktabs}
\DeclareMathOperator{\tr}{tr}

\newcommand{\Slash}[1]{{\ooalign{\hfil/\hfil\crcr$#1$}}}
\numberwithin{equation}{section}

\usepackage{url}

\begin{document}

\title{Axial $U(1)$ anomaly in a gravitational field via the gradient flow}

\author{%
\name{\fname{Okuto} \surname{Morikawa}}{1}
and
\name{\fname{Hiroshi} \surname{Suzuki}}{1,\ast}
}

\address{%
\affil{1}{Department of Physics, Kyushu University
744 Motooka, Nishi-ku, Fukuoka, 819-0395, Japan}
\email{hsuzuki@phys.kyushu-u.ac.jp}
}

\date{\today}

\begin{abstract}
A regularization-independent universal formula for the energy--momentum tensor
in gauge theory in the flat spacetime can be written down by employing the
so-called Yang--Mills gradient flow. We examine a possible use of the formula
in the calculation of the axial $U(1)$ anomaly in a gravitational field, the
anomaly first obtained by Toshiei Kimura [Prog.\ Theor.\ Phys.\  {\bf 42}, 1191
(1969)]. As a general argument indicates, the formula reproduces the correct
non-local structure of the
(axial $U(1)$ current)--(energy--momentum tensor)--(energy--momentum tensor)
triangle diagram in a way that is consistent with the axial $U(1)$ anomaly. On
the other hand, the formula does not automatically reproduce the general
coordinate (or translation) Ward--Takahashi relation, requiring corrections by
local counterterms. This analysis thus illustrates the fact that the universal
formula as it stands can be used only in on-shell correlation functions, in
which the energy--momentum tensor does not coincide with other composite
operators in coordinate space.
\end{abstract}
\subjectindex{B31, B32, B38}
\maketitle

\section{Introduction}
\label{sec:1}
Almost half a century ago, just nine months after the appearance of two seminal
papers on the axial $U(1)$ anomaly in an electromagnetic
field~\cite{Adler:1969gk,Bell:1969ts}, Kimura noticed in a lesser-known but
remarkable paper~\cite{Kumura:1969wj} that a similar anomalous non-conservation
of the axial vector current also occurs in a gravitational field. His result
was
\begin{equation}
   D^\alpha\left\langle\Bar{\psi}(x)\gamma_\alpha\gamma_5\psi(x)\right\rangle
   -2m_0\left\langle\Bar{\psi}(x)\gamma_5\psi(x)\right\rangle
   =\frac{1}{384\pi^2}\epsilon_{\mu\nu\rho\sigma}
   R^{\mu\nu\lambda\tau}(x)R^{\rho\sigma}{}_{\lambda\tau}(x),
\label{eq:(1.1)}
\end{equation}
where $m_0$ is the bare mass of the fermion and $R^{\mu\nu\rho\sigma}(x)$ is
the Riemann curvature. This axial $U(1)$ anomaly, also obtained
in~Refs.~\cite{Delbourgo:1972xb,Eguchi:1976db} (see also
Refs.~\cite{AlvarezGaume:1983wp,Fujikawa:1986hk}), was the first example of the
quantum anomaly related to the gravitational interaction, a subject that was to
be extensively explored somewhat later in a wider
context~\cite{AlvarezGaume:1983ig,Green:1984sg}.

Recently, by employing the so-called Yang--Mills gradient
flow~\cite{Narayanan:2006rf,Luscher:2009eq,Luscher:2010iy,Luscher:2011bx,
Luscher:2013cpa} (see~Refs.~\cite{Luscher:2013vga,Ramos:2015dla} for reviews)
and the small flow time expansion~\cite{Luscher:2011bx}, a
regularization-independent universal formula for the energy--momentum tensor in
gauge theory in the flat spacetime has been constructed~\cite{Suzuki:2013gza,
Makino:2014taa} (see also~Ref.~\cite{Suzuki:2016ytc} for a review); the formula
is then applied to the computation of thermodynamic quantities in lattice
QCD~\cite{Asakawa:2013laa,Taniguchi:2016ofw,Kitazawa:2016dsl,Ejiri:2017wgd,
Kitazawa:2017qab,Kanaya:2017cpp,Taniguchi:2017ibr}. References~\cite{%
DelDebbio:2013zaa,Fodor:2014cpa,Kikuchi:2014rla,Makino:2014sta,Makino:2014cxa,
Aoki:2014dxa,Suzuki:2015fka,Monahan:2015lha,Endo:2015iea,Suzuki:2015bqa,
Ramos:2015baa,Capponi:2015ahp,Datta:2015bzm,Christensen:2016wdo,
Fujikawa:2016qis,Hieda:2016lly,Kamata:2016any,Taniguchi:2016tjc,
Capponi:2016yjz,Hieda:2017sqq,Husung:2017qjz,Eller:2018yje} represent a
partial list of developments relating to the gradient flow.

In this paper, we examine a possible use of the universal formula in the
calculation of the axial $U(1)$ anomaly~\eqref{eq:(1.1)}; we will obtain
Eq.~\eqref{eq:(1.1)} by expansion around the flat spacetime. Precisely
speaking, the anomaly is a clash between the axial $U(1)$ Ward--Takahashi (WT)
relation and the general coordinate (and the local Lorentz) WT relation. A
general argument given in~Ref.~\cite{Kumura:1969wj}, which is analogous to the
argument in~Ref.~\cite{Adler:1969gk}, shows that the anomaly~\eqref{eq:(1.1)}
is independent of the adopted regularization as long as the regularization is
physically sensible and one imposes the general coordinate WT relation; the
structure~\eqref{eq:(1.1)} is robust in this sense.

In what follows, we will observe that the universal formula does not
automatically reproduce the correct WT relation associated with the general
coordinate (or translation in the flat spacetime) WT relation. The resulting
correlation functions, however, can be modified by adding appropriate
\emph{local terms\/} so that the translation WT relation holds. Then, as the
general argument implies, we have~Eq.~\eqref{eq:(1.1)}. This shows that the
universal formula reproduces the correct \emph{non-local structure\/} of the
(axial $U(1)$ current)--(energy--momentum tensor)--(energy--momentum tensor)
triangle diagram in a way that is consistent with the axial $U(1)$ anomaly.
This is expected without any calculation from the construction of the universal
formula~\cite{Suzuki:2013gza,Makino:2014taa}, but to check this point
explicitly is certainly assuring. On the other hand, this analysis illustrates
that the universal formula as it stands can be used only in on-shell
correlation functions, in which the energy--momentum tensor does not coincide
with other composite operators in coordinate space, because it does not
automatically reproduce the translation WT relation when operators coincide.
How to remedy this point in (a generalization of) the universal formula is a
forthcoming challenge.

This paper is organized as follows. In Sect.~\ref{sec:2}, we summarize the
naively expected form of the axial $U(1)$ and the general coordinate (or
translation) WT relations in the flat spacetime limit. The breaking of these
relations is regarded as the quantum anomaly. In~Sect.~\ref{sec:3}, using the
universal formulas for the energy--momentum tensor of the Dirac
fermion~\cite{Makino:2014taa} and the axial $U(1)$
current~\cite{Endo:2015iea,Hieda:2016lly}, we compute the total divergences
of the triangle diagram and extract the parts potentially corresponding to the
anomaly. Each of the axial $U(1)$ current and the energy--momentum tensors can
possess a different flow time, $t_1$, $t_2$, and~$t_3$; these are eventually
taken to be zero. We adopt a particular ordering of the limits, which turns out
to considerably simplify the calculation. We find that the translation WT
relation does not hold. In~Sect.~\ref{sec:4}, we seek an appropriate local term
added to the triangle diagram, which restores the translation WT relation.
Although our analysis here is quite analogous to that
of~Ref.~\cite{Endo:2015iea} on the triangle anomaly in gauge theory, partially
due to the fact that the translation WT relation also contains some two-point
functions, the analysis is much more complicated. Finally,
in~Sect.~\ref{sec:5}, by adding appropriate local terms, we obtain the
expansion of~Eq.~\eqref{eq:(1.1)} around the flat spacetime.
Section~\ref{sec:6} is devoted to the conclusion.

\section{Naively expected form of Ward--Takahashi relations}
\label{sec:2}
We consider the Dirac fermion in the curved spacetime with a Euclidean
signature. The curved space indices are denoted by Greek letters while the
local Lorentz indices are denoted by Latin letters. Letting $e_\mu^a(x)$ be the
vierbein, the raising and lowering of the former indices are done by the
metric $g_{\mu\nu}(x)\equiv\delta_{ab}e_\mu^a(x)e_\nu^b(x)$ and its inverse
matrix~$g^{\mu\nu}(x)$; those of the latter indices are, on the other hand, done
by the Kronecker deltas, $\delta_{ab}$ and~$\delta^{ab}$.

The action of the Dirac fermion in the curved spacetime is given by
\begin{equation}
   S=\int d^4x\,e(x)\Bar{\psi}(x)
   \left(\frac{1}{2}\overleftrightarrow{\Slash{D}}+m_0\right)\psi(x),
\label{eq:(2.1)}
\end{equation}
where $e(x)\equiv\det e_\mu^a(x)$,
$\overleftrightarrow{\Slash{D}}\equiv\Slash{D}-\overleftarrow{\Slash{D}}$,
and
\begin{align}
   \Slash{D}&\equiv e_a^\mu(x)\gamma^a
   \left[\partial_\mu+\frac{1}{4}\omega_\mu^{bc}(x)\sigma_{bc}\right]
   \equiv\gamma^\mu(x)D_\mu,
\label{eq:(2.2)}
\\
   \overleftarrow{\Slash{D}}&\equiv
   \left[\overleftarrow{\partial}_\mu
   -\frac{1}{4}\omega_\mu^{bc}(x)\sigma_{bc}\right]e_a^\mu(x)\gamma^a
   \equiv\overleftarrow{D}_\mu\gamma^\mu(x).
\label{eq:(2.3)}
\end{align}
$\gamma^a$ is the Dirac matrix satisfying $\{\gamma^a,\gamma^b\}=2\delta^{ab}$
and~$\sigma_{ab}\equiv\frac{1}{2}[\gamma_a,\gamma_b]$. $\omega_\mu^{ab}(x)$ is
the spin connection, which is defined by
\begin{align}
   \omega_\mu^{ab}(x)&\equiv\frac{1}{2}e^{a\rho}(x)e^{b\sigma}(x)
   \left[C_{\rho\sigma\mu}(x)-C_{\sigma\rho\mu}(x)-C_{\mu\rho\sigma}(x)\right],
\label{eq:(2.4)}
\\
   C_{\mu\rho\sigma}(x)&\equiv
   e_\mu^a(x)\left[\partial_\rho e_{a\sigma}(x)-\partial_\sigma e_{a\rho}(x)
   \right].
\label{eq:(2.5)}
\end{align}

The coupling of the fermion to a gravitational field is given by the
energy--momentum tensor:
\begin{align}
   T_{\mu\nu}(x)&\equiv\frac{1}{e}e_\mu^a(x)\frac{\delta}{\delta e^{\nu a}(x)}S
\notag\\
   &=\frac{1}{2}\Bar{\psi}(x)\gamma_\mu\overleftrightarrow{D}_\nu\psi(x)
   -g_{\mu\nu}\Bar{\psi}(x)
   \left(\frac{1}{2}\overleftrightarrow{\Slash{D}}+m_0\right)\psi(x)
   +\frac{1}{4}\epsilon_{\mu\nu\rho\sigma}D^\rho
   \left[\Bar{\psi}(x)\gamma_5\gamma^\sigma\psi(x)\right]
\notag\\
   &=T_{\mu\nu}^{\text{sym.}}(x)+T_{\mu\nu}^{\text{anti-sym.}}(x),
\label{eq:(2.6)}
\end{align}
where we have defined
\begin{align}
   T_{\mu\nu}^{\text{sym.}}(x)
   &\equiv\frac{1}{4}\Bar{\psi}(x)
   \left(\gamma_\mu\overleftrightarrow{D}_\nu
   +\gamma_\nu\overleftrightarrow{D}_\mu\right)\psi(x)
   -g_{\mu\nu}\Bar{\psi}(x)
   \left(\frac{1}{2}\overleftrightarrow{\Slash{D}}+m_0\right)\psi(x)
\label{eq:(2.7)}
\\
   T_{\mu\nu}^{\text{anti-sym.}}(x)
   &\equiv
   \frac{1}{4}\Bar{\psi}(x)\left[
   \sigma_{\mu\nu}(x)(\Slash{D}+m_0)
   +(\overleftarrow{\Slash{D}}-m_0)\sigma_{\mu\nu}(x)
   \right]\psi(x),
\label{eq:(2.8)}
\end{align}
where $\sigma_{\mu\nu}(x)\equiv e_\mu^a(x)e_\nu^b(x)\sigma_{ab}$. We note that
the anti-symmetric part of the energy--momentum
tensor~$T_{\mu\nu}^{\text{anti-sym.}}(x)$ is proportional to the equation of
motion of the fermion.

Now, in order to determine the precise form of the quantum anomalies, it is
crucial to clearly recognize the form of naively expected WT relations. For
simplicity, we consider the massless fermion $m_0=0$ in what follows.

We start from the WT relation associated with the axial $U(1)$ symmetry. For
this, we take the correlation function
\begin{equation}
   \left\langle T_{\mu\nu}^{\text{sym.}}(y)
   T_{\rho\sigma}^{\text{sym.}}(z)\right\rangle
   \equiv\int d\mu\,
   T_{\mu\nu}^{\text{sym.}}(y)T_{\rho\sigma}^{\text{sym.}}(z)\,e^{-S},
\label{eq:(2.9)}
\end{equation}
where $d\mu$ denotes the functional integration measure for the fermion field,
and make the change of integration variables in the form of a localized axial
$U(1)$ transformation:\footnote{The chiral matrix~$\gamma_5$ is
defined by $\gamma_5\equiv%
\frac{1}{4!}\epsilon_{abcd}\gamma^a\gamma^b\gamma^c\gamma^d$ by the totally
anti-symmetric tensor~$\epsilon_{abcd}$ being normalized as~$\epsilon_{0123}=1$.}
\begin{equation}
   \delta\psi(x)=i\theta(x)\gamma_5\psi(x),\qquad
   \delta\Bar{\psi}(x)=i\theta(x)\Bar{\psi}(x)\gamma_5.
\label{eq:(2.10)}
\end{equation}
Noting that the action~\eqref{eq:(2.1)} changes under this change of
variables as
\begin{equation}
   \delta S=-i\int d^4x\,e(x)\theta(x)D^\alpha j_{5\alpha}(x),\qquad
   j_{5\alpha}(x)\equiv\Bar{\psi}(x)\gamma_\alpha\gamma_5\psi(x),
\label{eq:(2.11)}
\end{equation}
and considering the flat spacetime limit $e_\mu^a(x)\to\delta_\mu^a$, neglecting
a possible breaking of the symmetry associated with the regularization, we
have the identity
\begin{align}
   &\partial_\alpha^x\left\langle j_{5\alpha}(x)
   T_{\mu\nu}^{\text{sym.}}(y)T_{\rho\sigma}^{\text{sym.}}(z)
   \right\rangle
\notag\\
   &\qquad{}
   +\partial_\alpha^x\delta(y-x)
   \left\langle
   \frac{1}{2}\Bar{\psi}(y)\gamma_5
   (\gamma_\mu\delta_{\nu\alpha}+\gamma_\nu\delta_{\mu\alpha}
   -2\delta_{\mu\nu}\gamma_\alpha)\psi(y)
   T_{\rho\sigma}^{\text{sym.}}(z)
   \right\rangle
\notag\\
   &\qquad{}
   +\partial_\alpha^x\delta(z-x)
   \left\langle
   T_{\mu\nu}^{\text{sym.}}(y)
   \frac{1}{2}\Bar{\psi}(z)\gamma_5
   (\gamma_\rho\delta_{\sigma\alpha}+\gamma_\sigma\delta_{\rho\alpha}
   -2\delta_{\rho\sigma}\gamma_\alpha)\psi(z)
   \right\rangle
\notag\\
   &=\partial_\alpha^x\left\langle j_{5\alpha}(x)
   T_{\mu\nu}^{\text{sym.}}(y)T_{\rho\sigma}^{\text{sym.}}(z)
   \right\rangle
\notag\\
   &=0,
\label{eq:(2.12)}
\end{align}
where the first equality follows from the covariance under the Lorentz and
parity transformations in the flat spacetime. The breaking of this
naively expected relation is thus regarded as a quantum anomaly.

Next, we consider the WT relation associated with the general coordinate
invariance and the local Lorentz symmetry. For this, we start with
\begin{equation}
   \left\langle j_{5\alpha}(x)T_{\rho\sigma}^{\text{sym.}}(z)\right\rangle
   \equiv\int d\mu\,j_{5\alpha}(x)T_{\rho\sigma}^{\text{sym.}}(z)\,e^{-S},
\label{eq:(2.13)}
\end{equation}
and consider the following form of the change of integration variables:
\begin{align}
   \delta\psi(x)
   &=\xi^\mu(x)\partial_\mu\psi(x)
   +\frac{1}{4}\xi^\mu(x)\omega_\mu^{ab}(x)\sigma_{ab}\psi(x)
   =\xi^\mu(x)D_\mu\psi(x),
\notag\\
   \delta\Bar{\psi}(x)
   &=\xi^\mu(x)\partial_\mu\Bar{\psi}(x)
   -\frac{1}{4}\xi^\mu(x)\Bar{\psi}(x)
   \sigma_{ab}\omega_\mu^{ab}(x)
   =\xi^\mu(x)D_\mu\Bar{\psi}(x).
\label{eq:(2.14)}
\end{align}
This is a particular combination of the general coordinate transformation and
the local Lorentz transformation. Under this change of integration variables,
from the fact that the action does not change if the vierbein is also changed
by the same set of transformations, we have
\begin{equation}
   \delta S=-\int d^4x\,e(x)\xi^\nu(x)D^\mu T_{\mu\nu}(x),
\label{eq:(2.15)}
\end{equation}
where the total energy--momentum tensor is given by~Eq.~\eqref{eq:(2.6)}.
Considering the flat spacetime limit, we thus have the identity
\begin{align}
   &\partial_\mu^y\left\langle j_{5\alpha}(x)T_{\mu\nu}^{\text{sym.}}(y)
   T_{\rho\sigma}^{\text{sym.}}(z)\right\rangle
\notag\\
   &\qquad{}
   +\partial_\mu^y\left\langle j_{5\alpha}(x)T_{\mu\nu}^{\text{anti-sym.}}(y)
   T_{\rho\sigma}^{\text{sym.}}(z)\right\rangle
\notag\\
   &\qquad{}
   +\delta(x-y)
   \partial_\nu^x\left\langle j_{5\alpha}(x)
   T_{\rho\sigma}^{\text{sym.}}(z)\right\rangle
\notag\\
   &\qquad{}
   +\delta(z-y)
   \partial_\nu^z\left\langle j_{5\alpha}(x)
   T_{\rho\sigma}^{\text{sym.}}(z)\right\rangle
\notag\\
   &\qquad{}
   +\partial_\beta^z\delta(z-y)
   \left\langle j_{5\alpha}(x)
   \frac{1}{4}\Bar{\psi}(z)
   (\gamma_\rho\delta_{\sigma\beta}+\gamma_\sigma\delta_{\rho\beta}
   -2\delta_{\rho\sigma}\gamma_\beta)\overleftrightarrow{\partial}_\nu\psi(z)
   \right\rangle
\notag\\
   &=\partial_\mu^y\left\langle j_{5\alpha}(x)T_{\mu\nu}^{\text{sym.}}(y)
   T_{\rho\sigma}^{\text{sym.}}(z)\right\rangle
\notag\\
   &\qquad{}
   +\partial_\mu^y\left\langle j_{5\alpha}(x)T_{\mu\nu}^{\text{anti-sym.}}(y)
   T_{\rho\sigma}^{\text{sym.}}(z)\right\rangle
\notag\\
   &\qquad{}
   +\partial_\beta^y\delta(y-z)
   \left\langle j_{5\alpha}(x)\mathcal{O}_{1\beta,\nu,\rho\sigma}(z)
   \right\rangle
\notag\\
   &=0,
\label{eq:(2.16)}
\end{align}
where again the first equality follows from the covariance under the Lorentz
and parity transformations. We have introduced the combination
\begin{equation}
   \mathcal{O}_{1\beta,\nu,\rho\sigma}(x)
   \equiv
   -\frac{1}{4}\Bar{\psi}(x)
   (\gamma_\rho\delta_{\sigma\beta}+\gamma_\sigma\delta_{\rho\beta}
   -2\delta_{\rho\sigma}\gamma_\beta)\overleftrightarrow{\partial}_\nu\psi(x).
\label{eq:(2.17)}
\end{equation}

On the other hand, by considering the change of integration variables of the
form of the local Lorentz transformation,
\begin{equation}
   \delta\psi(x)
   =-\frac{1}{4}\theta^{ab}(x)\sigma_{ab}\psi(x),\qquad
   \delta\Bar{\psi}(x)
   =\frac{1}{4}\Bar{\psi}(x)\sigma_{ab}\theta^{ab}(x),
\label{eq:(2.18)}
\end{equation}
in~Eq.~\eqref{eq:(2.13)}, we have the following identity in the flat spacetime
limit:
\begin{align}
   &\left\langle j_{5\alpha}(x)T_{\mu\nu}^{\text{anti-sym.}}(y)
   T_{\rho\sigma}^{\text{sym.}}(z)\right\rangle
\notag\\
   &=-\delta(x-y)
   \left\langle\frac{1}{4}
   \Bar{\psi}(x)[\gamma_\alpha\gamma_5,\sigma_{\mu\nu}]\psi(x)
   T_{\rho\sigma}^{\text{sym.}}(z)\right\rangle
\notag\\
   &\qquad{}
   -\delta(z-y)
   \left\langle j_{5\alpha}(x)
   \frac{1}{16}
   \Bar{\psi}(z)
   [\gamma_\rho\overleftrightarrow{\partial}_\sigma
   +\gamma_\sigma\overleftrightarrow{\partial}_\rho
   -2\delta_{\rho\sigma}\overleftrightarrow{\Slash{\partial}},\sigma_{\mu\nu}]
   \psi(z)
   \right\rangle
\notag\\
   &\qquad{}
   -\partial_\beta^z\delta(z-y)
   \left\langle j_{5\alpha}(x)
   \frac{1}{16}\Bar{\psi}(z)
   \{\gamma_\rho\delta_{\sigma\beta}+\gamma_\sigma\delta_{\rho\beta}
   -2\delta_{\rho\sigma}\gamma_\beta,\sigma_{\mu\nu}\}
   \psi(z)
   \right\rangle
\notag\\
   &=\delta(z-y)
   \left\langle j_{5\alpha}(x)\mathcal{O}_{2\mu\nu,\rho\sigma}(z)
   \right\rangle
\notag\\
   &\qquad{}
   +\partial_\beta^y\delta(z-y)
   \left\langle j_{5\alpha}(x)\mathcal{O}_{3\beta,\mu\nu,\rho\sigma}(z)
   \right\rangle,
\label{eq:(2.19)}
\end{align}
where the last equality again follows from the covariance under the Lorentz
and parity transformations and we have defined
\begin{align}
   \mathcal{O}_{2\mu\nu,\rho\sigma}(x)
   &\equiv-\frac{1}{16}
   \Bar{\psi}(x)
   [\gamma_\rho\overleftrightarrow{\partial}_\sigma
   +\gamma_\sigma\overleftrightarrow{\partial}_\rho
   -2\delta_{\rho\sigma}\overleftrightarrow{\Slash{\partial}},\sigma_{\mu\nu}]
   \psi(x),
\label{eq:(2.20)}
\\
   \mathcal{O}_{3\beta,\mu\nu,\rho\sigma}(x)
   &\equiv
   \frac{1}{16}\Bar{\psi}(x)
   \{\gamma_\rho\delta_{\sigma\beta}+\gamma_\sigma\delta_{\rho\beta}
   -2\delta_{\rho\sigma}\gamma_\beta,\sigma_{\mu\nu}\}
   \psi(x).
\label{eq:(2.21)}
\end{align}

Thus, combining Eqs.~\eqref{eq:(2.16)} and~\eqref{eq:(2.19)}, we have the
relation in the flat spacetime limit:
\begin{align}
   &\partial_\mu^y\left\langle j_{5\alpha}(x)T_{\mu\nu}^{\text{sym.}}(y)
   T_{\rho\sigma}^{\text{sym.}}(z)\right\rangle
\notag\\
   &\qquad{}
   +\partial_\beta^y\delta(y-z)
   \left\langle j_{5\alpha}(x)
   \mathcal{O}_{1\beta,\nu,\rho\sigma}(z)
   \right\rangle
\notag\\
   &\qquad{}
   +\partial_\mu^y\delta(y-z)
   \left\langle j_{5\alpha}(x)
   \mathcal{O}_{2\mu\nu,\rho\sigma}(z)
   \right\rangle
\notag\\
   &\qquad{}
   +\partial_\mu^y\partial_\beta^y\delta(y-z)
   \left\langle j_{5\alpha}(x)
   \mathcal{O}_{3\beta,\mu\nu,\rho\sigma}(z)
   \right\rangle
\notag\\
   &=0.
\label{eq:(2.22)}
\end{align}

Equation~\eqref{eq:(2.22)} is the naively expected form of the WT relation
associated with the general coordinate invariance and the local Lorentz
symmetry (in the flat spacetime limit). Thus, the breaking of this relation
should be regarded as a quantum anomaly. It can be confirmed that one can
directly derive the WT relation~\eqref{eq:(2.22)} only by using the
translational invariance in the flat spacetime (see Appendix~\ref{sec:A}). The
last three two-point functions including $\mathcal{O}_1$, $\mathcal{O}_2$,
and~$\mathcal{O}_3$ play a crucial role in the following analysis of the
anomaly. The contribution of such ``two-sided diagrams'' in addition to the
triangle diagram, which have no analogue in the axial $U(1)$ anomaly in gauge
theory, has of course already been noted in~Ref.~\cite{Kumura:1969wj} (through
a somewhat different derivation from ours).

\section{Computation of anomalies}
\label{sec:3}
\subsection{Definition of the three-point function}
\label{sec:3.1}
Now, for the vector-like gauge theory in the flat spacetime, we know
representations of the axial vector current~\cite{Endo:2015iea,Hieda:2016lly}
and the symmetric energy--momentum tensor~\cite{Suzuki:2013gza,Asakawa:2013laa,
Makino:2014taa} by the small flow time limit of flowed fields. In the zeroth
order in the gauge coupling, the representations are rather trivial:
\begin{align}
   j_{5\alpha}(t,x)&\equiv\Bar{\chi}(t,x)\gamma_\alpha\gamma_5\chi(t,x),
\label{eq:(3.1)}
\\
   T_{\mu\nu}^{\text{sym.}}(t,x)&\equiv\frac{1}{4}\Bar{\chi}(t,x)
   \left(\gamma_\mu\overleftrightarrow{\partial}_\nu
   +\gamma_\nu\overleftrightarrow{\partial}_\mu
   -2\delta_{\mu\nu}\overleftrightarrow{\Slash{\partial}}
   \right)\chi(t,x),
\label{eq:(3.2)}
\end{align}
where $\chi(t,x)$ and~$\Bar{\chi}(t,x)$ are flowed fermion fields and
eventually we have to take the small flow time limit $t\to0$ in the correlation
functions. Then, using the tree-level propagator of the flowed fermion
field~\cite{Luscher:2013cpa},\footnote{Throughout this paper, we use the
abbreviation
\begin{equation}
   \int_p\equiv\int\frac{d^4p}{(2\pi)^4}.
\label{eq:(3.3)}
\end{equation}}
\begin{equation}
   \left\langle\chi(t,x)\Bar{\chi}(s,y)\right\rangle_0
   =\int_p\,e^{ip(x-y)}\frac{e^{-(t+s)p^2}}{i\Slash{p}},
\label{eq:(3.4)}
\end{equation}
we have
\begin{align}
   &\left\langle j_{5\alpha}(x)
   T_{\mu\nu}^{\text{sym.}}(y)T_{\rho\sigma}^{\text{sym.}}(z)\right\rangle
\notag\\
   &\equiv\lim_{t_1\equiv t\to0}\lim_{t_2=t_3\to0}
   \left\langle j_{5\alpha}(t_1,x)
   T_{\mu\nu}^{\text{sym.}}(t_2,y)T_{\rho\sigma}^{\text{sym.}}(t_3,z)\right\rangle
\notag\\
   &=\lim_{t_1\equiv t\to0}\lim_{t_2=t_3\to0}
   \int_{p,q,k}\,e^{ip(x-y)}e^{iq(y-z)}e^{ik(z-x)}
   e^{-(t_1+t_2)p^2}e^{-(t_2+t_3)q^2}e^{-(t_3+t_1)k^2}
\notag\\
   &\qquad{}
   \times
   \frac{i}{16}\biggl(
   \tr\biggl\{
   \gamma_\alpha\gamma_5\frac{1}{\Slash{p}}
   \left[\gamma_\mu(p+q)_\nu+\gamma_\nu(p+q)_\mu
   -2\delta_{\mu\nu}(\Slash{p}+\Slash{q})\right]
   \frac{1}{\Slash{q}}
\notag\\
   &\qquad\qquad\qquad\qquad\qquad{}\times
   \left[\gamma_\rho(q+k)_\sigma+\gamma_\sigma(q+k)_\rho
   -2\delta_{\rho\sigma}(\Slash{q}+\Slash{k})\right]
   \frac{1}{\Slash{k}}
   \biggr\}
\notag\\
   &\qquad\qquad{}
   -\tr\biggl\{
   \gamma_\alpha\gamma_5\frac{1}{\Slash{k}}
   \left[\gamma_\rho(q+k)_\sigma+\gamma_\sigma(q+k)_\rho
   -2\delta_{\rho\sigma}(\Slash{q}+\Slash{k})\right]
   \frac{1}{\Slash{q}}
\notag\\
   &\qquad\qquad\qquad\qquad\qquad{}\times
   \left[\gamma_\mu(p+q)_\nu+\gamma_\nu(p+q)_\mu
   -2\delta_{\mu\nu}(\Slash{p}+\Slash{q})\right]
   \frac{1}{\Slash{p}}
   \biggr\}
   \biggr)
\notag\\
   &=\lim_{t\to0}
   \int_{p,q,k}\,e^{ip(x-y)}e^{iq(y-z)}e^{ik(z-x)}
   e^{-tp^2}e^{-tk^2}
\notag\\
   &\qquad{}
   \times
   \frac{i}{16}\biggl(
   \tr\biggl\{
   \gamma_\alpha\gamma_5\frac{1}{\Slash{p}}
   \left[\gamma_\mu(p+q)_\nu+\gamma_\nu(p+q)_\mu
   -2\delta_{\mu\nu}(\Slash{p}+\Slash{q})\right]
   \frac{1}{\Slash{q}}
\notag\\
   &\qquad\qquad\qquad\qquad\qquad{}\times
   \left[\gamma_\rho(q+k)_\sigma+\gamma_\sigma(q+k)_\rho
   -2\delta_{\rho\sigma}(\Slash{q}+\Slash{k})\right]
   \frac{1}{\Slash{k}}
   \biggr\}
\notag\\
   &\qquad\qquad{}
   -\tr\biggl\{
   \gamma_\alpha\gamma_5\frac{1}{\Slash{k}}
   \left[\gamma_\rho(q+k)_\sigma+\gamma_\sigma(q+k)_\rho
   -2\delta_{\rho\sigma}(\Slash{q}+\Slash{k})\right]
   \frac{1}{\Slash{q}}
\notag\\
   &\qquad\qquad\qquad\qquad\qquad{}\times
   \left[\gamma_\mu(p+q)_\nu+\gamma_\nu(p+q)_\mu
   -2\delta_{\mu\nu}(\Slash{p}+\Slash{q})\right]
   \frac{1}{\Slash{p}}
   \biggr\}
   \biggr).
\label{eq:(3.5)}
\end{align}
In this definition, we have adopted a particular ordering of the small time
limit; we first set $t_2=t_3\to0$ and then $t_1\equiv t\to0$. Because of the
Gaussian damping factors $e^{-(t_1+t_2)p^2}$, $e^{-(t_2+t_3)q^2}$,
and~$e^{-(t_3+t_1)k^2}$ in the first definition, the momentum integration is
absolutely convergent as long as $t_1\equiv t>0$; we can thus trivially take
the first limit $t_2=t_3\to0$ inside the momentum integration. It turns out
that this particular ordering considerably simplifies the actual calculation of
the anomalies below.\footnote{This simplification should be related to the fact
that the energy--momentum tensor induces the correct translation on composite
operators of the flowed fields with non-zero flow times (in our present case
$j_{5\alpha}(t,x)$), a fact emphasized in this context
in~Refs.~\cite{DelDebbio:2013zaa,Capponi:2015ahp,Capponi:2016yjz}.} We also
note that the expression~\eqref{eq:(3.5)} does not require further
regularization; in other words, Eq.~\eqref{eq:(3.5)} is independent of the
adopted regularization in the limit in which the regulator is sent to infinity.
This shows the universality of the representations~\eqref{eq:(3.1)}
and~\eqref{eq:(3.2)}, although this finiteness is trivial in the present
zeroth-order perturbation theory in the gauge coupling.

\subsection{Anomaly in the axial WT relation}
\label{sec:3.2}
We are primarily interested in the anomalous divergence of the axial vector
current, the breaking of the WT relation~\eqref{eq:(2.12)}. From our
definition~\eqref{eq:(3.5)}, after careful rearrangements, we find the identity
(omitting the symbol~$\lim_{t\to0}$)\footnote{We have noted that the spinor
trace with~$\gamma_5$ requires at least other four Dirac matrices.}
\begin{align}
   &\partial_\alpha^x\left\langle j_{5\alpha}(x)
   T_{\mu\nu}^{\text{sym.}}(y)T_{\rho\sigma}^{\text{sym.}}(z)
   \right\rangle
\notag\\
   &\qquad{}
   +\partial_\alpha^x\delta(y-x)
   \left\langle
   \frac{1}{2}\Bar{\psi}(y)\gamma_5
   (\gamma_\mu\delta_{\nu\alpha}+\gamma_\nu\delta_{\mu\alpha}
   -2\delta_{\mu\nu}\gamma_\alpha)\psi(y)
   T_{\rho\sigma}^{\text{sym.}}(z)
   \right\rangle
\notag\\
   &\qquad{}
   +\partial_\alpha^x\delta(z-x)
   \left\langle
   T_{\mu\nu}^{\text{sym.}}(y)
   \frac{1}{2}\Bar{\psi}(z)\gamma_5
   (\gamma_\rho\delta_{\sigma\alpha}+\gamma_\sigma\delta_{\rho\alpha}
   -2\delta_{\rho\sigma}\gamma_\alpha)\psi(z)
   \right\rangle
\notag\\
   &=\partial_\alpha^x\mathop\text{F.T.}e^{-tp^2}e^{-tk^2}
   \frac{-i}{8}
   \tr\left\{
   \gamma_5(
   \gamma_\mu\delta_{\nu\alpha}+\gamma_\nu\delta_{\mu\alpha}
   -2\delta_{\mu\nu}\gamma_\alpha)
   \frac{1}{\Slash{q}}
   \left[
   \gamma_\rho(q+k)_\sigma+\gamma_\sigma(q+k)_\rho
   \right]
   \frac{1}{\Slash{k}}\right\}
\notag\\
   &\qquad{}
   +\partial_\alpha^x\mathop\text{F.T.}e^{-tp^2}e^{-tk^2}
   \frac{i}{8}
   \tr\left\{
   \gamma_5
   \frac{1}{\Slash{p}}
   \left[
   \gamma_\mu(p+q)_\nu+\gamma_\nu(p+q)_\mu
   \right]
   \frac{1}{\Slash{q}} 
   (\gamma_\rho\delta_{\sigma\alpha}+\gamma_\sigma\delta_{\rho\alpha}
   -2\delta_{\rho\sigma}\gamma_\alpha)
 \right\}
\notag\\
   &\qquad{}
   +\mathop\text{F.T.}e^{-tp^2}e^{-tk^2}
   \frac{1}{8}
   \tr\left\{
   \gamma_5\left[
   \gamma_\mu(q+k)_\nu+\gamma_\nu(q+k)_\mu\right]
   \frac{1}{\Slash{q}}
   \left[\gamma_\rho(q+k)_\sigma+\gamma_\sigma(q+k)_\rho\right]
   \frac{1}{\Slash{k}}\right\}
\notag\\
   &\qquad{}
   +\mathop\text{F.T.}e^{-tp^2}e^{-tk^2}
   \frac{1}{8}
   \tr\left\{
   \gamma_5\frac{1}{\Slash{p}}
   \left[\gamma_\mu(p+q)_\nu+\gamma_\nu(p+q)_\mu\right]
   \frac{1}{\Slash{q}}
   \left[\gamma_\rho(p+q)_\sigma+\gamma_\sigma(p+q)_\rho\right]
   \right\},
\label{eq:(3.6)}
\end{align}
where $\text{F.T.}$ denotes the Fourier transformation:
\begin{equation}
   \text{F.T.}\equiv\int_{p,q,k}\,e^{ip(x-y)}e^{iq(y-z)}e^{ik(z-x)}\times.
\label{eq:(3.7)}
\end{equation}
On the left-hand side of~Eq.~\eqref{eq:(3.6)}, the two-point functions have
been defined by
\begin{align}
   &\left\langle
   \frac{1}{2}\Bar{\psi}(y)\gamma_5
   (\gamma_\mu\delta_{\nu\alpha}+\gamma_\nu\delta_{\mu\alpha}
   -2\delta_{\mu\nu}\gamma_\alpha)\psi(y)
   T_{\rho\sigma}^{\text{sym.}}(z)
   \right\rangle
\notag\\
   &\equiv\int_{q,k}\,e^{i(q-k)(y-z)}e^{-tq^2}e^{-tk^2}
\notag\\
   &\qquad\times\frac{i}{8}
   \tr\left\{
   \gamma_5(
   \gamma_\mu\delta_{\nu\alpha}+\gamma_\nu\delta_{\mu\alpha}
   -2\delta_{\mu\nu}\gamma_\alpha)
   \frac{1}{\Slash{q}}
   \left[
   \gamma_\rho(q+k)_\sigma+\gamma_\sigma(q+k)_\rho
   \right]
   \frac{1}{\Slash{k}}\right\}
   =0,
\label{eq:(3.8)}
\\
   &\left\langle
   T_{\mu\nu}^{\text{sym.}}(y)
   \frac{1}{2}\Bar{\psi}(z)\gamma_5
   (\gamma_\rho\delta_{\sigma\alpha}+\gamma_\sigma\delta_{\rho\alpha}
   -2\delta_{\rho\sigma}\gamma_\alpha)\psi(z)
   \right\rangle
\notag\\
   &\equiv
   \int_{p,q}\,e^{i(-p+q)(y-z)}e^{-tp^2}e^{-tq^2}
\notag\\
   &\qquad{}\times
   \frac{-i}{8}
   \tr\left\{
   \gamma_5
   \frac{1}{\Slash{p}}
   \left[
   \gamma_\mu(p+q)_\nu+\gamma_\nu(p+q)_\mu
   \right]
   \frac{1}{\Slash{q}} 
   (\gamma_\rho\delta_{\sigma\alpha}+\gamma_\sigma\delta_{\rho\alpha}
   -2\delta_{\rho\sigma}\gamma_\alpha)
   \right\}
   =0.
\label{eq:(3.9)}
\end{align}
These regularized two-point correlation functions identically vanish, as should
be the case from the Lorentz and parity covariance.

In deriving Eq.~\eqref{eq:(3.6)}, we first apply $\partial_\alpha^x$
to~Eq.~\eqref{eq:(3.5)}. In the integrand, this produces the
factor~$\Slash{p}-\Slash{k}$; each term of this is canceled by
$1/\Slash{p}$~and~$1/\Slash{k}$. We then use the identities $p+q=(p-k)+(q+k)$
and~$q+k=(k-p)+(p+q)$ and express the momentum $(p-k)_\alpha$ by the
derivative~$-i\partial_\alpha^x$. These manipulations give rise to the
right-hand side of~Eq.~\eqref{eq:(3.6)}. The last two terms on the left-hand
side of~Eq.~\eqref{eq:(3.6)} are simply zero, as noted above; the inclusion of
those terms, however, clearly shows the correspondence to the naively expected
axial WT relation~\eqref{eq:(2.12)}.

Thus, comparing Eq.~\eqref{eq:(3.6)} and~Eq.~\eqref{eq:(2.12)}, we find that
the anomalous breaking of the axial symmetry is given by the right-hand side
of~Eq.~\eqref{eq:(3.6)}. We note that in~Eq.~\eqref{eq:(3.6)} if Gaussian
factors such as $e^{-tp^2}$ and~$e^{-tk^2}$ are simply unity (i.e., if we could
naively set $t\to0$ before the momentum integration), then the right-hand side
identically vanishes. The fact is that there are Gaussian factors and they give
rise to a non-vanishing result. After a straightforward calculation in the
$t\to0$ limit, we find
\begin{align}
   &\partial_\alpha^x\left\langle j_{5\alpha}(x)
   T_{\mu\nu}^{\text{sym.}}(y)T_{\rho\sigma}^{\text{sym.}}(z)
   \right\rangle
\notag\\
   &=\int_{p,q}\,e^{ip(x-y)}e^{iq(x-z)}
   \frac{1}{(4\pi)^2}
\notag\\
   &\qquad{}
   \times
   \left\{\frac{1}{12}
   \epsilon_{\mu\rho\beta\gamma}p_\beta q_\gamma
   \left[
   q_\nu p_\sigma
   +\delta_{\nu\sigma}\left(
   -\frac{1}{t}
   +p^2
   +pq
   +q^2\right)
   \right]
   +(\mu\leftrightarrow\nu,\rho\leftrightarrow\sigma)
   \right\}.
\label{eq:(3.10)}
\end{align}

\subsection{Anomaly in the translation WT relation}
\label{sec:3.3}
Next, we investigate the anomalous breaking of the translation WT
relation~\eqref{eq:(2.22)}. From~Eq.~\eqref{eq:(3.5)}, after careful
rearrangements by using the relation
\begin{align}
   \gamma_\mu(p+q)_\nu+\gamma_\nu(p+q)_\mu
   &=2\gamma_\mu(p+q)_\nu+\Slash{p}\sigma_{\mu\nu}-\sigma_{\mu\nu}\Slash{q}
   -\frac{1}{2}\gamma_\mu\gamma_\nu(\Slash{p}-\Slash{q})
   +\frac{1}{2}(\Slash{p}-\Slash{q})\gamma_\nu\gamma_\mu
\notag\\
   &=2\gamma_\mu(p+q)_\nu-\sigma_{\mu\nu}\Slash{p}+\Slash{q}\sigma_{\mu\nu}
   +\frac{1}{2}\gamma_\mu\gamma_\nu(\Slash{p}-\Slash{q})
   -\frac{1}{2}(\Slash{p}-\Slash{q})\gamma_\nu\gamma_\mu,
\label{eq:(3.11)}
\end{align}
we have the identity\footnote{We have again noted that the spinor trace
with~$\gamma_5$ requires at least other four Dirac matrices.}
\begin{align}
   &\partial_\mu^y\left\langle j_{5\alpha}(x)T_{\mu\nu}^{\text{sym.}}(y)
   T_{\rho\sigma}^{\text{sym.}}(z)\right\rangle
\notag\\
   &\qquad{}
   +\delta(x-y)
   \partial_\nu^x\left\langle j_{5\alpha}(x)
   T_{\rho\sigma}^{\text{sym.}}(z)\right\rangle
\notag\\
   &\qquad{}
   +\delta(y-z)
   \partial_\nu^z\left\langle j_{5\alpha}(x)
   T_{\rho\sigma}^{\text{sym.}}(z)\right\rangle
\notag\\
   &\qquad{}
   -\partial_\mu^y\delta(x-y)
   \left\langle\frac{1}{4}
   \Bar{\psi}(x)[\gamma_\alpha\gamma_5,\sigma_{\mu\nu}]\psi(x)
   T_{\rho\sigma}^{\text{sym.}}(z)\right\rangle
\notag\\
   &\qquad{}
   +\partial_\beta^y\delta(y-z)
   \left\langle j_{5\alpha}(x)
   \mathcal{O}_{1\beta,\nu,\rho\sigma}(z)
   \right\rangle
\notag\\
   &\qquad{}
   +\partial_\mu^y\delta(y-z)
   \left\langle j_{5\alpha}(x)
   \mathcal{O}_{2\mu\nu,\rho\sigma}(z)
   \right\rangle
\notag\\
   &\qquad{}
   +\partial_\mu^y\partial_\beta^y\delta(y-z)
   \left\langle j_{5\alpha}(x)
   \mathcal{O}_{3\beta,\mu\nu,\rho\sigma}(z)
   \right\rangle
\notag\\
   &=(\partial_\nu^x+\partial_\nu^y)\mathop\text{F.T.}
   e^{-tp^2}e^{-tk^2}
   \frac{-i}{4}\tr\left\{
   \gamma_\alpha\gamma_5\frac{1}{\Slash{q}}
   \left[
   \gamma_\rho (q+k)_\sigma+\gamma_\sigma(q+k)_\rho\right]
   \frac{1}{\Slash{k}}\right\}
\notag\\
   &\qquad{}
   -\partial_\mu^y\mathop\text{F.T.}
   e^{-tp^2}e^{-tk^2}
\notag\\
   &\qquad\qquad{}
   \times\biggl(\frac{-i}{16}\tr\left\{
   \gamma_\alpha\gamma_5\sigma_{\mu\nu}\frac{1}{\Slash{q}}
   \left[
   \gamma_\rho(q+k)_\sigma+\gamma_\sigma(q+k)_\rho
   -2\delta_{\rho\sigma}(\Slash{q}+\Slash{k})\right]
   \frac{1}{\Slash{k}}\right\}
\notag\\
   &\qquad\qquad\qquad{}+
   \frac{-i}{16}\tr\left\{
   \gamma_\alpha\gamma_5\frac{1}{\Slash{k}}
   \left[
   \gamma_\rho(q+k)_\sigma+\gamma_\sigma(q+k)_\rho
   -2\delta_{\rho\sigma}(\Slash{q}+\Slash{k})\right]
   \frac{1}{\Slash{q}}\sigma_{\mu\nu}\right\}\biggr)
\notag\\
   &\qquad{}
   +\mathop\text{F.T.}
   e^{-tp^2}e^{-tk^2}
   \frac{1}{4}\tr\left\{
   \gamma_\alpha\gamma_5(q+k)_\nu\frac{1}{\Slash{q}}
   \left[
   \gamma_\rho(q+k)_\sigma+\gamma_\sigma(q+k)_\rho\right]
   \frac{1}{\Slash{k}}\right\}
\notag\\
   &\qquad{}
   +\mathop\text{F.T.}
   e^{-tp^2}e^{-tk^2}
   \frac{-1}{4}\tr\left\{
   \gamma_\alpha\gamma_5(p+k)_\nu\frac{1}{\Slash{p}}
   \left[
   \gamma_\rho(p+k)_\sigma+\gamma_\sigma(p+k)_\rho\right]
   \frac{1}{\Slash{k}}\right\}
\notag\\
   &\qquad{}
   +\partial_\beta^y\partial_\nu^x\mathop\text{F.T.}
   e^{-tp^2}e^{-tk^2}
   \frac{1}{4}\tr\left[
   \gamma_\alpha\gamma_5\frac{1}{\Slash{p}}
   \left(
   \gamma_\rho\delta_{\sigma\beta}+\gamma_\sigma\delta_{\rho\beta}
   -2\delta_{\rho\sigma}\gamma_\beta\right)
   \frac{1}{\Slash{k}}\right].
\label{eq:(3.12)}
\end{align}

In this expression, the two-point functions on the left-hand side have been
defined by
\begin{align}
   &\partial_\nu^x\left\langle j_{5\alpha}(x)
   T_{\rho\sigma}^{\text{sym.}}(z)\right\rangle
\notag\\
   &\equiv
   \partial_\nu^x\int_{q,k}\,e^{i(q-k)(x-z)}
   e^{-tq^2}e^{-tk^2}
   \frac{-i}{4}\tr\left\{
   \gamma_\alpha\gamma_5\frac{1}{\Slash{q}}
   \left[
   \gamma_\rho (q+k)_\sigma+\gamma_\sigma(q+k)_\rho\right]
   \frac{1}{\Slash{k}}\right\}
   =0,
\label{eq:(3.13)}
\\
   &\partial_\nu^z\left\langle j_{5\alpha}(x)
   T_{\rho\sigma}^{\text{sym.}}(z)\right\rangle
\notag\\
   &\equiv
   \partial_\nu^z
   \int_{p,k}\,e^{i(p-k)(x-z)}
   e^{-tp^2}e^{-tk^2}
   \frac{-i}{4}\tr\left\{
   \gamma_\alpha\gamma_5\frac{1}{\Slash{p}}
   \left[
   \gamma_\rho (p+k)_\sigma+\gamma_\sigma(p+k)_\rho\right]
   \frac{1}{\Slash{k}}\right\}
   =0,
\label{eq:(3.14)}
\\
   &\left\langle\frac{1}{4}
   \Bar{\psi}(x)[\gamma_\alpha\gamma_5,\sigma_{\mu\nu}]\psi(x)
   T_{\rho\sigma}^{\text{sym.}}(z)\right\rangle
\notag\\
   &\equiv
   \int_{q,k}\,e^{i(q-k)(x-z)}
   e^{-tq^2}e^{-tk^2}
   \frac{-i}{16}\tr\left\{
   [\gamma_\alpha\gamma_5,\sigma_{\mu\nu}]\frac{1}{\Slash{q}}
   \left[
   \gamma_\rho(q+k)_\sigma+\gamma_\sigma(q+k)_\rho
   -2\delta_{\rho\sigma}(\Slash{q}+\Slash{k})\right]
   \frac{1}{\Slash{k}}\right\}
\notag\\
   &=0,
\label{eq:(3.15)}
\end{align}
and
\begin{align}
   &\left\langle j_{5\alpha}(x)
   \mathcal{O}_{1\beta,\nu,\rho\sigma}(z)
   \right\rangle
\notag\\
   &\equiv
   \int_{p,k}\,e^{i(p-k)(x-z)}
   e^{-tp^2}e^{-tk^2}
   \frac{1}{4}\tr\left[
   \gamma_\alpha\gamma_5\frac{1}{\Slash{p}}
   \left(
   \gamma_\rho\delta_{\sigma\beta}+\gamma_\sigma\delta_{\rho\beta}
   -2\delta_{\rho\sigma}\gamma_\beta\right)i(p+k)_\nu
   \frac{1}{\Slash{k}}\right],
\label{eq:(3.16)}
\\
   &\left\langle j_{5\alpha}(x)
   \mathcal{O}_{2\mu\nu,\rho\sigma}(z)
   \right\rangle
\notag\\
   &\equiv
   \int_{p,k}\,e^{i(p-k)(x-z)}
   e^{-tp^2}e^{-tk^2}
   \frac{-i}{16}\tr\left[
   \gamma_\alpha\gamma_5\frac{1}{\Slash{k}}
   \left[
   \gamma_\rho(p+k)_\sigma+\gamma_\sigma(p+k)_\rho
   -2\delta_{\rho\sigma}(\Slash{p}+\Slash{k}),\sigma_{\mu\nu}\right]
   \frac{1}{\Slash{p}}\right],
\label{eq:(3.17)}
\\
   &\left\langle j_{5\alpha}(x)
   \mathcal{O}_{3\beta,\mu\nu,\rho\sigma}(z)
   \right\rangle
\notag\\
   &\equiv
   \int_{p,k}\,e^{i(p-k)(x-z)}
   e^{-tp^2}e^{-tk^2}
   \frac{1}{16}\tr\left(
   \gamma_\alpha\gamma_5\frac{1}{\Slash{p}}
   \left\{
   \gamma_\rho\delta_{\sigma\beta}+\gamma_\sigma\delta_{\rho\beta}
   -2\delta_{\rho\sigma}\gamma_\beta,\sigma_{\mu\nu}\right\}
   \frac{1}{\Slash{k}}\right).
\label{eq:(3.18)}
\end{align}
The two-point functions in~Eqs.~\eqref{eq:(3.13)}--\eqref{eq:(3.15)}
identically vanish as should be the case from the Lorentz and parity
covariance.

On the right-hand side of~Eq.~\eqref{eq:(3.12)}, the last two lines change sign
under the change of integration variables, $p\to-k$ and~$k\to-p$. Thus, those
two lines identically vanish. The other three terms do not vanish and after a
tedious calculation in the limit~$t\to0$, we have
\begin{align}
   &\partial_\mu^y\left\langle j_{5\alpha}(x)T_{\mu\nu}^{\text{sym.}}(y)
   T_{\rho\sigma}^{\text{sym.}}(z)\right\rangle
\notag\\
   &\qquad{}
   +\partial_\beta^y\delta(y-z)
   \left\langle j_{5\alpha}(x)
   \mathcal{O}_{1\beta,\nu,\rho\sigma}(z)
   \right\rangle
\notag\\
   &\qquad{}
   +\partial_\mu^y\delta(y-z)
   \left\langle j_{5\alpha}(x)
   \mathcal{O}_{2\mu\nu,\rho\sigma}(z)
   \right\rangle
\notag\\
   &\qquad{}
   +\partial_\mu^y\partial_\beta^y\delta(y-z)
   \left\langle j_{5\alpha}(x)
   \mathcal{O}_{3\beta,\mu\nu,\rho\sigma}(z)
   \right\rangle
\notag\\
   &=\int_{p,q}\,e^{ip(x-y)}e^{iq(x-z)}\frac{1}{(4\pi)^2}
\notag\\
   &\qquad{}
   \times
   \biggl\{
   p_\beta q_\gamma\epsilon_{\alpha\beta\gamma\nu}
   \left(
   \frac{1}{12}p_\rho p_\sigma+\frac{1}{12}p_\rho q_\sigma
   +\frac{1}{12}q_\rho p_\sigma-\frac{1}{8}pq\delta_{\rho\sigma}
   \right)
\notag\\
   &\qquad\qquad{}
   +p_\beta q_\gamma\epsilon_{\alpha\beta\gamma\rho}
   \left[
   -\frac{1}{12}q_\nu p_\sigma+\frac{1}{24}q_\nu q_\sigma
   +\delta_{\nu\sigma}\left(
   \frac{1}{12t}-\frac{1}{16}p^2-\frac{1}{12}pq-\frac{5}{48}q^2\right)
   \right]
   +(\rho\leftrightarrow\sigma)
\notag\\
   &\qquad\qquad{}
   +p_\beta\epsilon_{\alpha\beta\nu\rho}
   \left[
   p_\sigma
   \left(-\frac{1}{24t}+\frac{1}{48}p^2+\frac{1}{12}pq+\frac{5}{48}q^2\right)
   +q_\sigma\left(-\frac{1}{24}pq\right)\right]
   +(\rho\leftrightarrow\sigma)
\notag\\
   &\qquad\qquad{}
   +q_\beta\epsilon_{\alpha\beta\nu\rho}
   \left[
   p_\sigma
   \left(-\frac{1}{12t}+\frac{1}{48}p^2+\frac{1}{12}pq+\frac{5}{48}q^2\right)
   +q_\sigma\left(-\frac{1}{24}pq\right)\right]
   +(\rho\leftrightarrow\sigma)
   \biggr\}.
\label{eq:(3.19)}
\end{align}

\subsection{Axial anomaly in the two-point functions}
\label{sec:3.4}
For the subsequent analysis, we still need to know possible anomalous
breakings of the axial WT relations for the two-point
functions~\eqref{eq:(3.16)}--\eqref{eq:(3.18)}. From the structure of the
fermion bi-linear operators, $\mathcal{O}_1$ (Eq.~\eqref{eq:(2.17)}),
$\mathcal{O}_2$ (Eq.~\eqref{eq:(2.20)}), and~$\mathcal{O}_3$
(Eq.~\eqref{eq:(2.21)}), we would expect, as the axial $U(1)$ WT relations,
$\partial_\alpha^x\langle j_{5\alpha}(x)\mathcal{O}_{1\beta,\nu,\rho\sigma}(z)
\rangle=
\partial_\alpha^x\langle j_{5\alpha}(x)\mathcal{O}_{2\mu\nu,\rho\sigma}(z)
\rangle=
\partial_\alpha^x\langle j_{5\alpha}(x)\mathcal{O}_{3\beta,\mu\nu,\rho\sigma}(z)
\rangle=0$. It turns out that our definition~\eqref{eq:(3.18)} is not
compatible with this expectation and
\begin{align}
   &\partial_\alpha^x\left\langle j_{5\alpha}(x)
   \mathcal{O}_{1\beta,\nu,\rho\sigma}(z)
   \right\rangle=0,
\label{eq:(3.20)}
\\
   &\partial_\alpha^x\left\langle j_{5\alpha}(x)
   \mathcal{O}_{2\mu\nu,\rho\sigma}(z)
   \right\rangle=0,
\label{eq:(3.21)}
\\
   &\partial_\alpha^x\left\langle j_{5\alpha}(x)
   \mathcal{O}_{3\beta,\mu\nu,\rho\sigma}(z)
   \right\rangle
\notag\\
   &=i\int_q\,e^{iq(x-z)}\frac{1}{(4\pi)^2}
   (\epsilon_{\rho\mu\nu\gamma}\delta_{\sigma\beta}
   +\epsilon_{\sigma\mu\nu\gamma}\delta_{\rho\beta}
   -2\delta_{\rho\sigma}\epsilon_{\beta\mu\nu\gamma})q_\gamma
   \left(\frac{1}{8t}-\frac{1}{12}q^2\right).
\label{eq:(3.22)}
\end{align}
As we will see, however, the axial anomaly in the two-point
function~\eqref{eq:(3.22)} can be removed by adding an appropriate local term
to the two-point function~\eqref{eq:(3.18)}.

\section{Local counterterms}
\label{sec:4}
Now, the anomalous breaking of the axial WT relation in~Eq.~\eqref{eq:(3.10)}
would have intrinsic meaning only when we require the validity of the
translation WT relation~\eqref{eq:(2.22)}. That is, we still have the freedom
to modify the local part of the three-point correlation
function~\eqref{eq:(3.5)} by adding a ``local counterterm''
$\mathcal{C}_{\alpha,\mu\nu,\rho\sigma}(x,y,z)$. In the momentum space, it must be
a cubic polynomial of external momenta. We see that the general form of the
counterterm that is consistent with the symmetric structure of the three-point
function~\eqref{eq:(3.5)} is given by
\begin{align}
   &\mathcal{C}_{\alpha,\mu\nu,\rho\sigma}(x,y,z)
\notag\\
   &=\int_{p,q}\,e^{ip(x-y)}e^{iq(x-z)}\frac{i}{(4\pi)^2}
\notag\\
   &\qquad{}
   \times\bigl[
   \epsilon_{\alpha\mu\rho\beta}
   p_\beta\delta_{\nu\sigma}(c_0+c_1p^2+c_2pq+c_3q^2)
   -\epsilon_{\alpha\mu\rho\beta}q_\beta\delta_{\nu\sigma}
   (c_0+c_3p^2+c_2pq+c_1q^2)
\notag\\
   &\qquad\qquad{}
   +\epsilon_{\alpha\mu\rho\beta}p_\beta
   (d_1p_\nu p_\sigma+d_2p_\nu q_\sigma+d_3q_\nu p_\sigma+d_4q_\nu q_\sigma)
\notag\\
   &\qquad\qquad{}
   -\epsilon_{\alpha\mu\rho\beta}q_\beta
   (d_4p_\nu p_\sigma+d_2p_\nu q_\sigma+d_3q_\nu p_\sigma+d_1q_\nu q_\sigma)
\notag\\
   &\qquad\qquad{}
   +\epsilon_{\alpha\mu\beta\gamma}p_\beta q_\gamma
   (e_1p_\nu\delta_{\rho\sigma}+e_2p_\rho\delta_{\nu\sigma}
   +e_3q_\nu\delta_{\rho\sigma}+e_4q_\rho\delta_{\nu\sigma})
\notag\\
   &\qquad\qquad{}
   -\epsilon_{\alpha\rho\beta\gamma}p_\beta q_\gamma
   (e_3p_\sigma\delta_{\mu\nu}+e_4p_\mu\delta_{\nu\sigma}
   +e_1q_\sigma\delta_{\mu\nu}+e_2q_\mu\delta_{\nu\sigma})
\notag\\
   &\qquad\qquad{}
   +\epsilon_{\mu\rho\beta\gamma}p_\beta q_\gamma
   (f_1p_\alpha\delta_{\nu\sigma}+f_2p_\nu\delta_{\alpha\sigma}
   +f_3p_\sigma\delta_{\alpha\nu}
   +f_1q_\alpha\delta_{\nu\sigma}+f_3q_\nu\delta_{\alpha\sigma}
   +f_2q_\sigma\delta_{\alpha\nu})
\notag\\
   &\qquad\qquad{}
   +(\mu\leftrightarrow\nu,\rho\leftrightarrow\sigma)
   \bigr],
\label{eq:(4.1)}
\end{align}
where $c_i$, $d_i$, $e_i$, and~$f_i$ are constants. The basic idea is to choose
the coefficients $c_i$, $d_i$, $e_i$, and~$f_i$ so that the right-hand side
of~Eq.~\eqref{eq:(3.19)} vanishes after the addition $\langle j_{5\alpha}(x)
T_{\mu\nu}^{\text{sym.}}(y)T_{\rho\sigma}^{\text{sym.}}(z)\rangle+
\mathcal{C}_{\alpha,\mu\nu,\rho\sigma}(x,y,z)$. Then, to the axial
anomaly~\eqref{eq:(3.10)}, the counterterm contributes by
\begin{align}
   &\partial_\alpha^x\mathcal{C}_{\alpha,\mu\nu,\rho\sigma}(x,y,z)
\notag\\
   &=\int_{p,q}\,e^{ip(x-y)}e^{iq(x-z)}\frac{1}{(4\pi)^2}
   \epsilon_{\mu\rho\beta\gamma}p_\beta q_\gamma
\notag\\
   &\qquad{}
   \times\bigl\{
   (d_1+d_4-f_2-f_3)p_\nu p_\sigma+(2d_2-2f_2)p_\nu q_\sigma
\notag\\
   &\qquad\qquad{}
   +(2d_3-2f_3)q_\nu p_\sigma+(d_1+d_4-f_2-f_3)q_\nu q_\sigma
\notag\\
   &\qquad\qquad{}
   +\delta_{\nu\sigma}
   \left[2c_0+(c_1+c_3-f_1)p^2+(2c_2-2f_1)pq+(c_1+c_3-f_1)q^2\right]
   \bigr\}
\notag\\
   &\qquad\qquad{}
   +(\mu\leftrightarrow\nu,\rho\leftrightarrow\sigma).
\label{eq:(4.2)}
\end{align}

A complication arises, however, since we may also modify the two-point
functions~\eqref{eq:(3.16)}--\eqref{eq:(3.18)} appearing in the
relation~\eqref{eq:(3.19)} by adding local terms. We choose the counterterms
for the two-point functions such that the axial $U(1)$ WT relations hold for
the two-point functions. 

For the two-point function~\eqref{eq:(3.16)}, we thus require the validity of
the axial WT relation,
\begin{equation}
   \partial_\alpha^x
   \left[\left\langle
   j_{5\alpha}(x)\mathcal{O}_{1\beta,\nu,\rho\sigma}(z)
   \right\rangle
   +\mathcal{S}_{1\alpha,\beta,\nu,\rho\sigma}(x,z)\right]=0,
\label{eq:(4.3)}
\end{equation}
where $\mathcal{S}_{1\alpha,\beta,\nu,\rho\sigma}(x,z)$ is a local term.
Equation~\eqref{eq:(3.20)}, however, shows that there is no axial anomaly in this
two-point function and thus we should require
$\partial_\alpha^x\mathcal{S}_{1\alpha,\beta,\nu,\rho\sigma}(x,z)=0$. It turns out
that the most general form of such a local term is
\begin{align}
   &\mathcal{S}_{1\alpha,\beta,\nu,\rho\sigma}(x,z)
\notag\\
   &=\int_q\,e^{iq(x-z)}\frac{i}{(4\pi)^2}
\notag\\
   &\qquad{}
   \times
   \Bigl\{
   \epsilon_{\alpha\beta\nu\rho}q_\sigma
   \left[\Tilde{c}_0+(\Tilde{c}_1+\Tilde{c}_2)q^2\right]
   +(\rho\leftrightarrow\sigma)
\notag\\
   &\qquad\qquad{}
   +\epsilon_{\beta\nu\rho\gamma}q_\gamma
   \left[\delta_{\alpha\sigma}(\Tilde{c}_0+\Tilde{c}_1q^2)
   +\Tilde{c}_2q_\alpha q_\sigma\right]
   +(\rho\leftrightarrow\sigma)
\notag\\
   &\qquad\qquad{}
   +\epsilon_{\alpha\beta\nu\gamma}q_\gamma
   \left[\delta_{\rho\sigma}(\Tilde{d}_0+\Tilde{d}_1q^2)
   +\Tilde{d}_2q_\rho q_\sigma\right]
\notag\\
   &\qquad\qquad{}
   +\epsilon_{\alpha\beta\rho\gamma}q_\gamma
   \left[\delta_{\nu\sigma}(\Tilde{e}_0+\Tilde{e}_1q^2)
   +\Tilde{e}_2q_\nu q_\sigma\right]
   +(\rho\leftrightarrow\sigma)
\notag\\
   &\qquad\qquad{}
   +\epsilon_{\alpha\nu\rho\gamma}q_\gamma
   \left[\delta_{\beta\sigma}(\Tilde{f}_0+\Tilde{f}_1q^2)
   +\Tilde{f}_2q_\beta q_\sigma\right]
   +(\rho\leftrightarrow\sigma)
   \Bigr\},
\label{eq:(4.4)}
\end{align}
where $\Tilde{c}_i$, $\Tilde{d}_i$, $\Tilde{e}_i$, and~$\Tilde{f}_i$ are
constants.

Similarly, for the two-point function~\eqref{eq:(3.17)}, requiring
\begin{equation}
   \partial_\alpha^x\left[\left\langle
   j_{5\alpha}(x)\mathcal{O}_{2\mu\nu,\rho\sigma}(z)
   \right\rangle
   +\mathcal{S}_{2\alpha,\mu\nu,\rho\sigma}(x,z)\right]=0
\label{eq:(4.5)}
\end{equation}
implies $\partial_\alpha^x\mathcal{S}_{2\alpha,\mu\nu,\rho\sigma}(x,z)=0$
because of~Eq.~\eqref{eq:(3.21)} and the possible form of the counterterm is
given by
\begin{align}
   &\mathcal{S}_{2\alpha,\mu\nu,\rho\sigma}(x,z)
\notag\\
   &=\int_q\,e^{iq(x-z)}\frac{i}{(4\pi)^2}
\notag\\
   &\qquad{}\times
   \Bigl\{
   \epsilon_{\alpha\mu\nu\rho}q_\sigma
   \left[\Tilde{c}_0'+(\Tilde{c}_1'+\Tilde{c}_2')q^2\right]
   +(\rho\leftrightarrow\sigma)
\notag\\
   &\qquad\qquad{}
   +\epsilon_{\mu\nu\rho\beta}q_\beta
   \left[\delta_{\alpha\sigma}(\Tilde{c}_0'+\Tilde{c}_1'q^2)
   +\Tilde{c}_2'q_\alpha q_\sigma\right]
   +(\rho\leftrightarrow\sigma)
\notag\\
   &\qquad\qquad{}
   +\epsilon_{\alpha\mu\nu\beta}q_\beta
   \left[\delta_{\rho\sigma}(\Tilde{d}_0'+\Tilde{d}_1'q^2)
   +\Tilde{d}_2'q_\rho q_\sigma\right]
\notag\\
   &\qquad\qquad{}
   +\epsilon_{\alpha\mu\rho\beta}q_\beta
   \left[\delta_{\nu\sigma}(\Tilde{e}_0'+\Tilde{e}_1'q^2)
   +\Tilde{e}_2'q_\nu q_\sigma\right]
   -\epsilon_{\alpha\nu\rho\beta}q_\beta
   \left[\delta_{\mu\sigma}(\Tilde{e}_0'+\Tilde{e}_1'q^2)
   +\Tilde{e}_2'q_\mu q_\sigma\right]
\notag\\
   &\qquad\qquad\qquad{}
   +(\rho\leftrightarrow\sigma)
   \Bigr\},
\label{eq:(4.6)}
\end{align}
where $\Tilde{c}_i'$, $\Tilde{d}_i'$, and~$\Tilde{e}_i'$ are constants.

Finally, after some examination, we find that the most general form of the
counterterm for the function~\eqref{eq:(3.18)} is given by
\begin{align}
   &\mathcal{S}_{3\alpha,\beta,\mu\nu,\rho\sigma}(x,z)
\notag\\
   &=\int_q\,e^{iq(x-z)}\frac{1}{(4\pi)^2}
\notag\\
   &\qquad{}
   \times\bigl\{
   \epsilon_{\alpha\rho\mu\nu}
   (c_0'\delta_{\sigma\beta}+c_1'q^2\delta_{\sigma\beta}+c_2'q_\sigma q_\beta)
   +(\rho\leftrightarrow\sigma)
\notag\\
   &\qquad\qquad{}
   +\epsilon_{\alpha\beta\mu\nu}
   (d_0'\delta_{\rho\sigma}+d_1'q^2\delta_{\rho\sigma}+d_2'q_\rho q_\sigma)
\notag\\
   &\qquad\qquad{}
   +\left[
   \epsilon_{\alpha\beta\mu\rho}
   (e_0'\delta_{\nu\sigma}+e_1'q^2\delta_{\nu\sigma}+e_2'q_\nu q_\sigma)
   -\epsilon_{\alpha\beta\nu\rho}
   (e_0'\delta_{\mu\sigma}+e_1'q^2\delta_{\mu\sigma}+e_2'q_\mu q_\sigma)
   +(\rho\leftrightarrow\sigma)
   \right]
\notag\\
   &\qquad\qquad{}
   +\epsilon_{\rho\mu\nu\gamma}q_\gamma
   (f_1'q_\beta\delta_{\alpha\sigma}+f_2'q_\sigma\delta_{\alpha\beta}
   +f_3'q_\alpha\delta_{\beta\sigma})
   +(\rho\leftrightarrow\sigma)
\notag\\
   &\qquad\qquad{}
   +\epsilon_{\beta\mu\nu\gamma}q_\gamma
   (g_1'q_\rho\delta_{\alpha\sigma}+g_1'q_\sigma\delta_{\alpha\rho}
   +g_2'q_\alpha\delta_{\rho\sigma})
\notag\\
   &\qquad\qquad{}
   +\bigl[
   \epsilon_{\beta\mu\rho\gamma}q_\gamma
   (h_1'q_\nu\delta_{\alpha\sigma}+h_2'q_\sigma\delta_{\alpha\nu}
   +h_3'q_\alpha\delta_{\nu\sigma})
\notag\\
   &\qquad\qquad\qquad{}
   -\epsilon_{\beta\nu\rho\gamma}q_\gamma
   (h_1'q_\mu\delta_{\alpha\sigma}+h_2'q_\sigma\delta_{\alpha\mu}
   +h_3'q_\alpha\delta_{\mu\sigma})
   +(\rho\leftrightarrow\sigma)
   \bigr]
\notag\\
   &\qquad\qquad{}
   +\epsilon_{\alpha\mu\nu\gamma}q_\gamma
   (c_1''q_\rho\delta_{\beta\sigma}+c_1''q_\sigma\delta_{\beta\rho}
   +c_2''q_\beta\delta_{\rho\sigma})
\notag\\
   &\qquad\qquad{}
   +\left[
   \epsilon_{\alpha\beta\mu\gamma}q_\gamma
   (d_1''q_\rho\delta_{\nu\sigma}+d_1''q_\sigma\delta_{\nu\rho}
   +d_2''q_\nu\delta_{\rho\sigma})
   -\epsilon_{\alpha\beta\nu\gamma}q_\gamma
   (d_1''q_\rho\delta_{\mu\sigma}+d_1''q_\sigma\delta_{\mu\rho}
   +d_2'' q_\mu\delta_{\rho\sigma})
   \right]
\notag\\
   &\qquad\qquad{}
   +\epsilon_{\alpha\beta\rho\gamma}q_\gamma
   (e_1''q_\mu\delta_{\nu\sigma}-e_1''q_\nu\delta_{\mu\sigma})
   +(\rho\leftrightarrow\sigma)
\notag\\
   &\qquad\qquad{}
   +\bigl[
   \epsilon_{\alpha\mu\rho\gamma}q_\gamma
   (f_1''q_\beta\delta_{\nu\sigma}+f_2''q_\nu\delta_{\beta\sigma}
   +f_3'' q_\sigma\delta_{\beta\nu})
\notag\\
   &\qquad\qquad\qquad{}
   -\epsilon_{\alpha\nu\rho\gamma}q_\gamma
   (f_1''q_\beta\delta_{\mu\sigma}+f_2''q_\mu\delta_{\beta\sigma}
   +f_3''q_\sigma\delta_{\beta\mu})
   +(\rho\leftrightarrow\sigma)
   \bigr]
   \bigr\}.
\label{eq:(4.7)}
\end{align}
We choose the coefficients $c_0'$ etc.\ so that the addition
of~$\mathcal{S}_{1\alpha,\beta,\mu\nu,\rho\sigma}(x,z)$ to the two-point function
cancels the anomalous breaking~\eqref{eq:(3.22)}. That is, we require
\begin{equation}
   \partial_\alpha^x
   \left[\left\langle
   j_{5\alpha}(x)\mathcal{O}_{3\beta,\mu\nu,\rho\sigma}(z)
   \right\rangle
   +\mathcal{S}_{3\alpha,\beta,\mu\nu,\rho\sigma}(x,z)\right]=0.
\label{eq:(4.8)}
\end{equation}
This yields
\begin{align}
   c_0'&=\frac{1}{8t},&
   c_1'&=f_3'-\frac{1}{12},&
   c_2'&=f_1'+f_2',
\notag\\
   d_0'&=-\frac{1}{4t},&
   d_1'&=g_2'+\frac{1}{6},&
   d_2'&=2g_1',
\notag\\
   e_0'&=0,&
   e_1'&=h_3',&
   e_2'&=h_1'+h_2'.
\label{eq:(4.9)}
\end{align}

Now, we require that the translation WT relation~\eqref{eq:(2.22)} holds by
adding the above local terms to the correlation functions. That is, our
requirement is
\begin{align}
   &\partial_\mu^y
   \left[\left\langle j_{5\alpha}(x)T_{\mu\nu}^{\text{sym.}}(y)
   T_{\rho\sigma}^{\text{sym.}}(z)\right\rangle
   +\mathcal{C}_{\alpha,\mu\nu,\rho\sigma}(x,y,z)\right]
\notag\\
   &\qquad{}
   +\partial_\mu^y\partial_\beta^y\delta(y-z)
   \left[
   \left\langle j_{5\alpha}(x)
   \mathcal{O}_{3\beta,\mu\nu,\rho\sigma}(z)
   \right\rangle
   +\mathcal{S}_{1\alpha,\beta,\mu\nu,\rho\sigma}(x,z)
   \right]
\notag\\
   &\qquad{}
   +\partial_\beta^y\delta(y-z)
   \left[\left\langle j_{5\alpha}(x)
   \mathcal{O}_{1\beta,\nu,\rho\sigma}(z)
   \right\rangle
   +\mathcal{S}_{2\alpha,\beta,\nu,\rho\sigma}(x,z)
   \right]
\notag\\
   &\qquad{}
   +\partial_\mu^y\delta(y-z)
   \left[\left\langle j_{5\alpha}(x)
   \mathcal{O}_{2\mu\nu,\rho\sigma}(z)
   \right\rangle
   +\mathcal{S}_{3\alpha,\mu\nu,\rho\sigma}(x,z)
   \right]
\notag\\
   &=0.
\label{eq:(4.10)}
\end{align}
The resulting relations among the coefficients in the counterterms are
summarized in Appendix~\ref{sec:B}. From those relations, we see that some
coefficients are still left unfixed, but the coefficients in the
expression~\eqref{eq:(4.2)} are completely determined as
\begin{align}
   &d_1+d_4-f_2-f_3=0,
\\
   &2d_2-2f_2=0,
\\
   &2d_3-2f_3=\frac{1}{12},
\\
   &2c_0=\frac{1}{12t},
\\
   &c_1+c_3-f_1=-f_1=-\frac{1}{12},
\\
   &2c_2-2f_1=-\frac{1}{4}.
\end{align}
This gives
\begin{align}
   &\partial_\alpha^x\mathcal{C}_{\alpha,\mu\nu,\rho\sigma}(x,y,z)
 \notag\\
   &=\int_{p,q}\,e^{ip(x-y)}e^{iq(x-z)}\frac{1}{(4\pi)^2}
   \epsilon_{\mu\rho\beta\gamma}p_\beta q_\gamma
   \left\{
   \frac{1}{12}q_\nu p_\sigma
   +\delta_{\nu\sigma}
   \left[\frac{1}{12t}-\frac{1}{12}p^2-\frac{1}{4}pq-\frac{1}{12}q^2\right]
   \right\}
\notag\\
   &\qquad\qquad\qquad\qquad\qquad{}
   +(\mu\leftrightarrow\nu,\rho\leftrightarrow\sigma).
\label{eq:(4.17)}
\end{align}

\section{Final steps}
\label{sec:5}
We are now able to write down the axial $U(1)$ anomaly in the three-point
function~\eqref{eq:(3.5)} under the requirement of the translation WT
relation~\eqref{eq:(4.10)}; the latter requirement is accomplished by the
counterterm~\eqref{eq:(4.1)}. Then the axial $U(1)$ anomaly is given by the
sum of~Eqs.~\eqref{eq:(3.10)} and~\eqref{eq:(4.17)}, i.e.,
\begin{align}
   &\partial_\alpha^x
   \left[\left\langle
   j_{5\alpha}(x)
   T_{\mu\nu}^{\text{sym.}}(y)T_{\rho\sigma}^{\text{sym.}}(z)\right\rangle
   +\mathcal{C}_{\alpha,\mu\nu,\rho\sigma}(x,y,z)\right]
\notag\\
   &=\int_{p,q}\,e^{ip(x-y)}e^{iq(x-z)}\frac{1}{(4\pi)^2}
   \left[
   \frac{1}{6}\epsilon_{\mu\rho\beta\gamma}p_\beta q_\gamma
   \left(q_\nu p_\sigma-\delta_{\nu\sigma}pq\right)
   +(\mu\leftrightarrow\nu,\rho\leftrightarrow\sigma)
   \right].
\label{eq:(5.1)}
\end{align}
This is the most non-trivial result of this paper.

Going back to Eq.~\eqref{eq:(2.6)}, the energy--momentum tensor also has a part
that is anti-symmetric under the exchange of indices,
$T_{\mu\nu}^{\text{anti-sym.}}(x)$ (Eq.~\eqref{eq:(2.8)}). We may redefine the
energy--momentum tensor~$T_{\mu\nu}(x)$ by simply removing this anti-symmetric
part because $T_{\mu\nu}^{\text{anti-sym.}}(x)$ in~Eq.~\eqref{eq:(2.8)} is
proportional to the equation of motion; its effect on the correlation functions
must be at most local contact terms as the Schwinger--Dyson equation implies.
We can in fact corroborate this argument by explicit calculations by using some
regularizing prescription for~$T_{\mu\nu}^{\text{anti-sym.}}(x)$. Here, however,
we are content with the above argument and set
$T_{\mu\nu}(x)\to T_{\mu\nu}^{\text{sym.}}(x)$ in what follows.

We now re-express Eq.~\eqref{eq:(5.1)} as the anomalous divergence of the
axial $U(1)$ current in the curved spacetime. We expand
$D^\alpha\langle j_{5\alpha}(x)\rangle_g$, the divergence of the axial vector
current in the curved spacetime as the power series of the vierbein around the
flat spacetime:
\begin{align}
   &D^\alpha\left\langle j_{5\alpha}(x)\right\rangle_g
\notag\\
   &=D^\alpha\left\langle j_{5\alpha}(x)\right\rangle
   +\int d^4y\,\delta e^{\mu a}(y)\frac{\delta}{\delta e^{\mu a}(x)}
   D^\alpha\left\langle j_{5\alpha}(x)\right\rangle
\notag\\
   &\qquad{}
   +\frac{1}{2!}
   \int d^4y\,\delta e^{\mu a}(y)\int d^4z\,\delta e^{\nu b}(z)
   \frac{\delta}{\delta e^{\mu a}(y)}\frac{\delta}{\delta e^{\nu b}(z)}
   D^\alpha\left\langle j_{5\alpha}(x)\right\rangle+O(\delta e^3),
\label{eq:(5.2)}
\end{align}
where $\delta e^{\mu a}(x)\equiv e^{\mu a}(x)-\delta^{\mu a}$ and it is understood
that the right-hand side is evaluated in the flat spacetime with appropriate
local counterterms specified as above. Noting
\begin{equation}
   \left\langle j_{5\alpha}(x)\right\rangle=0,\qquad
   \left\langle j_{5\alpha}(x)T_{\mu\nu}^{\text{sym.}}(y)\right\rangle=0,\qquad
   \left\langle\partial_\alpha j_{5\alpha}(x)\frac{\delta}{\delta e^{\rho b}(z)}
   T_{\mu\nu}^{\text{sym.}}(y)\right\rangle=0,
\label{eq:(5.3)}
\end{equation}
as far as the regularization preserves the Lorentz and parity covariance, we
have
\begin{align}
   &D^\alpha\left\langle j_{5\alpha}(x)\right\rangle_g
\notag\\
   &=\frac{1}{2!}
   \int d^4y\,\delta e^{\mu a}(y)e_a^\rho(y)
   \int d^4z\,\delta e^{\nu b}(z)e_b^\sigma(z)
   \left\langle\partial_\alpha j_{5\alpha}(x)
   T_{\rho\mu}^{\text{sym.}}(y)T_{\sigma\nu}^{\text{sym.}}(z)\right\rangle
   +O(\delta e^3)
\notag\\
   &=\frac{1}{2!}
   \int d^4y\,\frac{1}{2}\delta g^{\mu\nu}(y)
   \int d^4z\,\frac{1}{2}\delta g^{\rho\sigma}(z)
   \partial_\alpha^x\left\langle j_{5\alpha}(x)
   T_{\mu\nu}^{\text{sym.}}(y)T_{\rho\sigma}^{\text{sym.}}(z)\right\rangle
   +O(\delta g^3),
\notag\\
   &=\frac{1}{(4\pi)^2}\frac{1}{12}
   \epsilon_{\mu\nu\rho\sigma}\partial_\mu\partial_\lambda\delta g_{\nu\tau}(x)
   \left[\partial_\rho\partial_\lambda\delta g_{\sigma\tau}(x)
   -\partial_\rho\partial_\tau\delta g_{\sigma\lambda}(x)\right]
   +O(\delta g^3),
\label{eq:(5.4)}
\end{align}
where we have used $[\delta/\delta e^{\nu a}(x)]S=%
e^\mu_a(x)T_{\mu\nu}^{\text{sym.}}(x)$, $\delta g^{\mu\nu}(x)=%
\delta e^{\mu a}(x)e_a^\nu(x)+e^{\mu a}(x)\delta e_a^\nu(x)$
and~Eq.~\eqref{eq:(5.1)} in the last equality. Comparing this with the
expansion of the curvature,\footnote{Our definition of the Riemann curvature is
$R^\alpha{}_{\rho\mu\nu}
\equiv\partial_\mu{\mit\Gamma}_{\rho\nu}^\alpha
-\partial_\nu{\mit\Gamma}_{\rho\mu}^\alpha
+{\mit\Gamma}_{\lambda\mu}^\alpha{\mit\Gamma}_{\rho\nu}^\lambda
-{\mit\Gamma}_{\lambda\nu}^\alpha{\mit\Gamma}_{\rho\mu}^\lambda$, where
${\mit\Gamma}_{\mu\nu}^\lambda
\equiv\frac{1}{2}g^{\lambda\rho}
(\partial_\mu g_{\nu\rho}+\partial_\nu g_{\mu\rho}-\partial_\rho g_{\mu\nu})$
is the Christoffel symbol.}
\begin{equation}
   \epsilon_{\mu\nu\rho\sigma}
   R^{\mu\nu\lambda\tau}(x)R^{\rho\sigma}{}_{\lambda\tau}(x)
   =2\epsilon_{\mu\nu\rho\sigma}\partial_\mu\partial_\lambda\delta g_{\nu\tau}(x)
   \left[\partial_\rho\partial_\lambda\delta g_{\sigma\tau}(x)
   -\partial_\rho\partial_\tau\delta g_{\sigma\lambda}(x)\right]
   +O(\delta g^3),
\end{equation}
we finally observe Eq.~\eqref{eq:(1.1)} for $m_0=0$.

\section{Conclusion}
\label{sec:6}
In this paper, we have examined a possible use of the universal formula for the
energy--momentum tensor in gauge theory in the flat spacetime through the
Yang--Mills gradient flow~\cite{Suzuki:2013gza,Makino:2014taa}. As a general
argument indicates, after choosing local counterterms appropriately so as to
restore the translation WT relation, we obtain the correct axial $U(1)$
anomaly in~Eq.~\eqref{eq:(1.1)} (in the flat space limit).

From the present analysis, we can learn the following feature of the universal
formula of the energy--momentum tensor. The universal formula is based on the
gradient flow and its small flow time expansion of~Ref.~\cite{Luscher:2011bx}.
The latter asserts that any composite operator of flowed fields as~$t\to0$ can
be expressed as an asymptotic series of renormalized composite operators of
unflowed fields with increasing mass dimensions. When two composite operators
of flowed fields collide in coordinate space to form another composite
operator, we have to consider the expansion in terms of another set of
renormalized composite operators of unflowed fields. Consequently, it is not
obvious what happens when the universal formula of the energy--momentum tensor
collides with other composite operators, such as the axial $U(1)$ current or
the energy--momentum tensor, in coordinate space. Our present analysis
illustrates that the formula in fact does not automatically fulfill the
translation WT relation precisely when the formula coincides with other
composite operators in coordinate space. On the other hand, our finding that
\emph{local\/} counterterms are sufficient to restore the translation WT
relation ensures the expectation that the formula fulfills the translation WT
relation when the energy--momentum tensor is in isolation in coordinate space;
for this case, the translation WT relation is simply the conservation law of
the energy--momentum tensor.

Thus, our analysis has revealed that the universal formula as it stands can be
used only in on-shell correlation functions (i.e., correlation functions in
which the energy--momentum tensor does not coincide with other composite
operators in coordinate space). The incorporate of this point into (a
generalization of) the universal formula is a forthcoming
challenge.\footnote{We would like to thank Shinya Aoki for an interesting
suggestion on this issue.} A related issue is the possible generalization of
the gradient flow to the curved spacetime. A possible generalization is
\begin{align}
   \partial_t B_\mu(t,x)&=g^{\nu\rho}(x)D_\nu G_{\rho,\mu}(t,x),&
   B_\mu(t=0,x)&=A_\mu(x),
\\
   \partial_t\chi(t,x)&=g^{\mu\nu}(x)D_\mu D_\nu\chi(t,x),&
   \chi(t=0,x)&=\psi(x),
\\
   \partial_t\Bar{\chi}(t,x)
   &=\Bar{\chi}(t,x)g^{\mu\nu}(x)\overleftarrow{D}_\mu\overleftarrow{D}_\mu,&
   \Bar{\chi}(t=0,x)&=\Bar{\psi}(x).
\end{align}
It then appears interesting to see whether this setup improves the covariance
under the general coordinate transformation and the restoration of the
associated WT relations for the energy--momentum tensor.

\section*{Acknowledgements}
This work was originally planned to be presented at the commemorative lecture
for the Yukawa--Kimura prize of 2017.
The work of H. S. is supported in part by a JSPS Grant-in-Aid for Scientific
Research Grant Number JP16H03982.

\appendix

\section{Derivation of the WT relation~\eqref{eq:(2.22)} in the flat spacetime}
\label{sec:A}
In the functional integral corresponding to the correlation
function~$\langle j_{5\alpha}(x)T_{\rho\sigma}^{\text{sym.}}(z)\rangle$ in the flat
spacetime, where
\begin{align}
   T_{\mu\nu}^{\text{sym.}}(x)
   \equiv\frac{1}{4}\Bar{\psi}(x)
   \left(\gamma_\mu\overleftrightarrow{\partial}_\nu
   +\gamma_\nu\overleftrightarrow{\partial}_\mu\right)\psi(x)
   -\delta_{\mu\nu}\Bar{\psi}(x)
   \left(\frac{1}{2}\overleftrightarrow{\Slash{\partial}}+m_0\right)\psi(x),
\label{eq:(A1)}
\end{align}
we consider the change of integration variables of the form of the localized
translation:
\begin{equation}
   \delta\psi(x)=\xi_\mu(x)\partial_\mu\psi(x),\qquad
   \delta\Bar{\psi}(x)=\xi_\mu(x)\partial_\mu\Bar{\psi}(x).
\label{eq:(A2)}
\end{equation}
Since the action in the flat spacetime
\begin{equation}
   S=\int d^4x\,\Bar{\psi}(x)
   \left(\frac{1}{2}\overleftrightarrow{\Slash{\partial}}+m_0\right)\psi(x)
\label{eq:(A3)}
\end{equation}
changes under Eq.~\eqref{eq:(A2)} as
\begin{equation}
   \delta S=-\int d^4x\,\xi_\nu(x)\partial_\mu T_{\mu\nu}^{\text{can.}}(x),
\label{eq:(A4)}
\end{equation}
where $T_{\mu\nu}^{\text{can.}}(x)$ is the \emph{canonical\/} energy--momentum
tensor, defined by
\begin{equation}
   T_{\mu\nu}^{\text{can.}}(x)
   \equiv\frac{1}{2}\Bar{\psi}(x)\gamma_\mu\overleftrightarrow{\partial}_\nu
   \psi(x)
   -\delta_{\mu\nu}\Bar{\psi}(x)
   \left(\frac{1}{2}\overleftrightarrow{\Slash{\partial}}+m_0\right)
   \psi(x),
\label{eq:(A5)}
\end{equation}
we have
\begin{align}
   &\partial_\mu^y\left\langle j_{5\alpha}(x)T_{\mu\nu}^{\text{can.}}(y)
   T_{\rho\sigma}^{\text{sym.}}(z)\right\rangle
\notag\\
   &\qquad{}
   +\delta(x-y)
   \partial_\nu^x\left\langle j_{5\alpha}(x)
   T_{\rho\sigma}^{\text{sym.}}(z)\right\rangle
\notag\\
   &\qquad{}
   +\delta(z-y)
   \partial_\nu^z\left\langle j_{5\alpha}(x)
   T_{\rho\sigma}^{\text{sym.}}(z)\right\rangle
\notag\\
   &\qquad{}
   +\partial_\beta^y\delta(y-z)
   \left\langle j_{5\alpha}(x)\mathcal{O}_{1\beta,\nu,\rho\sigma}(z)
   \right\rangle
\notag\\
   &=0,
\label{eq:(A6)}
\end{align}
where $\mathcal{O}_{1\beta,\nu,\rho\sigma}(x)$ is the combination defined
in~Eq.~\eqref{eq:(2.17)}.

We next note that the symmetric energy--momentum tensor~\eqref{eq:(A1)} and
the canonical energy--momentum tensor~\eqref{eq:(A5)} are related as
(this is the relation attributed to Belinfante and~Rosenfeld)
\begin{align}
   T_{\mu\nu}^{\text{sym.}}(x)
   &=T_{\mu\nu}^{\text{can.}}(x)
   -\frac{1}{4}
   \Bar{\psi}(x)\left[
   \sigma_{\mu\nu}(\Slash{\partial}+m_0)
   +(\overleftarrow{\Slash{\partial}}-m_0)\sigma_{\mu\nu}
   \right]\psi(x)
\notag\\
   &\qquad{}+\frac{1}{8}
   \partial_\rho
   \left[
   \Bar{\psi}(x)\left(
   \gamma_\mu\gamma_\nu\gamma_\rho-\gamma_\rho\gamma_\nu\gamma_\mu
   \right)\psi(x)
   \right].
\label{eq:(A7)}
\end{align}
On the right-hand side, since the last term has no total divergence with
respect to the index~$\mu$, it can be neglected in the following discussion.
The second term is proportional to the equations of motion and its insertion
in~$\langle j_{5\alpha}(x)T_{\rho\sigma}^{\text{sym.}}(z)\rangle$ can be determined
by the Schwinger--Dyson equation as
\begin{align}
   &\left\langle j_{5\alpha}(x)
   \left(-\frac{1}{4}\right)
   \Bar{\psi}(y)\left[
   \sigma_{\mu\nu}(\Slash{\partial}+m_0)
   +(\overleftarrow{\Slash{\partial}}-m_0)\sigma_{\mu\nu}
   \right]\psi(y)
   T_{\rho\sigma}^{\text{sym.}}(z)\right\rangle
\notag\\
   &=\delta(x-y)
   \left\langle\frac{1}{4}
   \Bar{\psi}(x)[\gamma_\alpha\gamma_5,\sigma_{\mu\nu}]\psi(x)
   T_{\rho\sigma}^{\text{sym.}}(z)\right\rangle
\notag\\
   &\qquad{}
   -\delta(z-y)
   \left\langle j_{5\alpha}(x)\mathcal{O}_{2\mu\nu,\rho\sigma}(z)
   \right\rangle
\notag\\
   &\qquad{}
   -\partial_\beta^y\delta(z-y)
   \left\langle j_{5\alpha}(x)\mathcal{O}_{3\beta,\mu\nu,\rho\sigma}(z)
   \right\rangle,
\label{eq:(A8)}
\end{align}
where the combinations $\mathcal{O}_{2\mu\nu,\rho\sigma}(x)$
and~$\mathcal{O}_{3\beta,\mu\nu,\rho\sigma}(x)$ are given
in~Eqs.~\eqref{eq:(2.20)} and~\eqref{eq:(2.21)}, respectively. Finally,
combining Eqs.~\eqref{eq:(A6)}, \eqref{eq:(A7)}, and~\eqref{eq:(A8)}, we have
Eq.~\eqref{eq:(2.22)}.

\section{Relations among counterterm coefficients}
\label{sec:B}
For the coefficients in~Eqs.~\eqref{eq:(4.1)}, \eqref{eq:(4.4)},
\eqref{eq:(4.6)}, and~\eqref{eq:(4.7)}, the requirements of the validity of
the WT relations~\eqref{eq:(4.8)} (i.e., Eq.~\eqref{eq:(4.9)})
and~\eqref{eq:(4.10)} yield the following relations:
\begin{align}
   c_0&=\frac{\Tilde{f}_0-\Tilde{e}_0'}{3}=\frac{1}{24t},
\\
   c_1&=-f_1''-f_2''+f_3''+\frac{1}{48},
\\
   c_2&=-\frac{1}{24},
\\
   c_3&=f_1''+f_2''-f_3''-\frac{1}{48}=-c_1,
\\
   d_1&=\Tilde{f}_2-\Tilde{e}_2'-\frac{1}{24},
\\
   d_2&=\Tilde{f}_2-\Tilde{e}_2'+f_3'',
\\
   d_3&=\Tilde{f}_2-\Tilde{e}_2'+f_3''+\frac{1}{24},
\\
   d_4&=\Tilde{f}_2-\Tilde{e}_2'+2f_3''+\frac{1}{24},
\\
   e_1&=f_3'',
\\
   e_2&=f_1''+f_2''-\frac{1}{24},
\\
   e_3&=f_3''+\frac{1}{16},
\\
   e_4&=f_1''+f_2''-\frac{1}{12},
\\
   f_1&=\frac{1}{12},
\\
   f_2&=\Tilde{f}_2-\Tilde{e}_2'+f_3'',
\\
   f_3&=\Tilde{f}_2-\Tilde{e}_2'+f_3'',
\end{align}
and
\begin{align}
   c_0'&=\Tilde{f}_0-\Tilde{e}_0'=\frac{1}{8t},
\\
   c_1'&=h_3',
\\
   c_2'&=h_1'+h_2'+2\Tilde{f}_2-2\Tilde{e}_2'+2f_3'',
\\
   d_0'&=-2(\Tilde{f}_0-\Tilde{e}_0')=-\frac{1}{4t},
\\
   d_1'&=g_2'+\frac{1}{6},
\\
   d_2'&=2g_1',
\\
   e_0'&=0,
\\
   e_1'&=h_3',
\\
   e_2'&=h_1'+h_2',
\\
   f_1'&=h_1'+\Tilde{f}_2-\Tilde{e}_2'+f_3'',
\\
   f_2'&=h_2'+\Tilde{f}_2-\Tilde{e}_2'+f_3'',
\\
   f_3'&=h_3'+\frac{1}{12},
\\
   c_1''&=d_1''-f_1''-f_2'',
\\
   c_2''&=d_2''-2 f_3'',
\\
   e_1''&=f_3''-f_1'',
\\
   \Tilde{c}_0&=-\Tilde{c}_0',
\\
   \Tilde{c}_1&=-\Tilde{c}_1',
\\
   \Tilde{c}_2&=-\Tilde{c}_2',
\\
   \Tilde{d}_0&=-\Tilde{d}_0',
\\
   \Tilde{d}_1&=-\Tilde{d}_1',
\\
   \Tilde{d}_2&=-\Tilde{d}_2',
\\
   \Tilde{e}_0&=\Tilde{f}_0-2\Tilde{e}_0'=-\Tilde{e}_0'+\frac{1}{8t},
\\
   \Tilde{e}_1&=-\Tilde{e}_1'-f_1''-f_2''+f_3''-\frac{1}{12},
\\
   \Tilde{e}_2&=\Tilde{f}_2-2\Tilde{e}_2',
\\
   \Tilde{f}_0&=\Tilde{e}_0'+\frac{1}{8t},
\\
   \Tilde{f}_1&=\Tilde{e}_1'-f_1''-f_2''+f_3''-\frac{1}{12}.
\end{align}


\begin{thebibliography}{00}

\bibitem{Adler:1969gk} 
  S.~L.~Adler,
  Phys.\ Rev.\  {\bf 177}, 2426 (1969).
  doi:10.1103/PhysRev.177.2426

\bibitem{Bell:1969ts} 
  J.~S.~Bell and R.~Jackiw,
  Nuovo Cim.\ A {\bf 60}, 47 (1969).
  doi:10.1007/BF02823296

\bibitem{Kumura:1969wj} 
  T.~Kimura,
  Prog.\ Theor.\ Phys.\  {\bf 42}, 1191 (1969).
  doi:10.1143/PTP.42.1191

\bibitem{Delbourgo:1972xb} 
  R.~Delbourgo and A.~Salam,
  Phys.\ Lett.\  {\bf 40B}, 381 (1972).
  doi:10.1016/0370-2693(72)90825-8

\bibitem{Eguchi:1976db} 
  T.~Eguchi and P.~G.~O.~Freund,
  Phys.\ Rev.\ Lett.\  {\bf 37}, 1251 (1976).
  doi:10.1103/PhysRevLett.37.1251

\bibitem{AlvarezGaume:1983wp} 
  L.~Alvarez-Gaum\'e,
  J.\ Phys.\ A {\bf 16}, 4177 (1983).
  doi:10.1088/0305-4470/16/18/018

\bibitem{Fujikawa:1986hk} 
  K.~Fujikawa, S.~Ojima and S.~Yajima,
  Phys.\ Rev.\ D {\bf 34}, 3223 (1986).
  doi:10.1103/PhysRevD.34.3223

\bibitem{AlvarezGaume:1983ig} 
  L.~Alvarez-Gaum\'e and E.~Witten,
  Nucl.\ Phys.\ B {\bf 234}, 269 (1984).
  doi:10.1016/0550-3213(84)90066-X

\bibitem{Green:1984sg} 
  M.~B.~Green and J.~H.~Schwarz,
  Phys.\ Lett.\  {\bf 149B}, 117 (1984).
  doi:10.1016/0370-2693(84)91565-X

\bibitem{Narayanan:2006rf} 
  R.~Narayanan and H.~Neuberger,
  JHEP {\bf 0603}, 064 (2006)
  doi:10.1088/1126-6708/2006/03/064
  [hep-th/0601210].

\bibitem{Luscher:2009eq} 
  M.~L\"uscher,
  Commun.\ Math.\ Phys.\  {\bf 293}, 899 (2010)
  doi:10.1007/s00220-009-0953-7
  [arXiv:0907.5491 [hep-lat]].

\bibitem{Luscher:2010iy} 
  M.~L\"uscher,
  JHEP {\bf 1008}, 071 (2010)
  Erratum: [JHEP {\bf 1403}, 092 (2014)]
  doi:10.1007/JHEP08(2010)071, 10.1007/JHEP03(2014)092
  [arXiv:1006.4518 [hep-lat]].

\bibitem{Luscher:2011bx} 
  M.~L\"uscher and P.~Weisz,
  JHEP {\bf 1102}, 051 (2011)
  doi:10.1007/JHEP02(2011)051
  [arXiv:1101.0963 [hep-th]].

\bibitem{Luscher:2013cpa} 
  M.~L\"uscher,
  JHEP {\bf 1304}, 123 (2013)
  doi:10.1007/JHEP04(2013)123
  [arXiv:1302.5246 [hep-lat]].

\bibitem{Luscher:2013vga} 
  M.~L\"uscher,
  PoS LATTICE {\bf 2013}, 016 (2014)
  [arXiv:1308.5598 [hep-lat]].

\bibitem{Ramos:2015dla} 
  A.~Ramos,
  PoS LATTICE {\bf 2014}, 017 (2015)
  [arXiv:1506.00118 [hep-lat]].

\bibitem{Suzuki:2013gza} 
  H.~Suzuki,
  PTEP {\bf 2013}, 083B03 (2013)
  Erratum: [PTEP {\bf 2015}, 079201 (2015)]
  doi:10.1093/ptep/ptt059, 10.1093/ptep/ptv094
  [arXiv:1304.0533 [hep-lat]].

\bibitem{Makino:2014taa} 
  H.~Makino and H.~Suzuki,
  PTEP {\bf 2014}, 063B02 (2014)
  Erratum: [PTEP {\bf 2015}, 079202 (2015)]
  doi:10.1093/ptep/ptu070, 10.1093/ptep/ptv095
  [arXiv:1403.4772 [hep-lat]].

\bibitem{Suzuki:2016ytc} 
  H.~Suzuki,
  PoS LATTICE {\bf 2016}, 002 (2017)
  [arXiv:1612.00210 [hep-lat]].

\bibitem{Asakawa:2013laa} 
  M.~Asakawa {\it et al.} [FlowQCD Collaboration],
  Phys.\ Rev.\ D {\bf 90}, no. 1, 011501 (2014)
  Erratum: [Phys.\ Rev.\ D {\bf 92}, no. 5, 059902 (2015)]
  doi:10.1103/PhysRevD.90.011501, 10.1103/PhysRevD.92.059902
  [arXiv:1312.7492 [hep-lat]].

\bibitem{Taniguchi:2016ofw} 
  Y.~Taniguchi, S.~Ejiri, R.~Iwami, K.~Kanaya, M.~Kitazawa, H.~Suzuki, T.~Umeda and N.~Wakabayashi,
  Phys.\ Rev.\ D {\bf 96}, no. 1, 014509 (2017)
  doi:10.1103/PhysRevD.96.014509
  [arXiv:1609.01417 [hep-lat]].

\bibitem{Kitazawa:2016dsl} 
  M.~Kitazawa, T.~Iritani, M.~Asakawa, T.~Hatsuda and H.~Suzuki,
  Phys.\ Rev.\ D {\bf 94}, no. 11, 114512 (2016)
  doi:10.1103/PhysRevD.94.114512
  [arXiv:1610.07810 [hep-lat]].

\bibitem{Ejiri:2017wgd} 
  S.~Ejiri {\it et al.},
  PoS LATTICE {\bf 2016}, 058 (2017)
  [arXiv:1701.08570 [hep-lat]].

\bibitem{Kitazawa:2017qab} 
  M.~Kitazawa, T.~Iritani, M.~Asakawa and T.~Hatsuda,
  Phys.\ Rev.\ D {\bf 96}, no. 11, 111502 (2017)
  doi:10.1103/PhysRevD.96.111502
  [arXiv:1708.01415 [hep-lat]].

\bibitem{Kanaya:2017cpp} 
  K.~Kanaya {\it et al.} [WHOT-QCD Collaboration],
  EPJ Web Conf.\  {\bf 175}, 07023 (2018)
  doi:10.1051/epjconf/201817507023
  [arXiv:1710.10015 [hep-lat]].

\bibitem{Taniguchi:2017ibr} 
  Y.~Taniguchi {\it et al.} [WHOT-QCD Collaboration],
  EPJ Web Conf.\  {\bf 175}, 07013 (2018)
  doi:10.1051/epjconf/201817507013
  [arXiv:1711.02262 [hep-lat]].

\bibitem{DelDebbio:2013zaa} 
  L.~Del Debbio, A.~Patella and A.~Rago,
  JHEP {\bf 1311}, 212 (2013)
  doi:10.1007/JHEP11(2013)212
  [arXiv:1306.1173 [hep-th]].

\bibitem{Fodor:2014cpa} 
  Z.~Fodor, K.~Holland, J.~Kuti, S.~Mondal, D.~Nogradi and C.~H.~Wong,
  JHEP {\bf 1409}, 018 (2014)
  doi:10.1007/JHEP09(2014)018
  [arXiv:1406.0827 [hep-lat]].

\bibitem{Kikuchi:2014rla} 
  K.~Kikuchi and T.~Onogi,
  JHEP {\bf 1411}, 094 (2014)
  doi:10.1007/JHEP11(2014)094
  [arXiv:1408.2185 [hep-th]].

\bibitem{Makino:2014sta} 
  H.~Makino and H.~Suzuki,
  PTEP {\bf 2015}, no. 3, 033B08 (2015)
  doi:10.1093/ptep/ptv028
  [arXiv:1410.7538 [hep-lat]].

\bibitem{Makino:2014cxa} 
  H.~Makino, F.~Sugino and H.~Suzuki,
  PTEP {\bf 2015}, no. 4, 043B07 (2015)
  doi:10.1093/ptep/ptv044
  [arXiv:1412.8218 [hep-lat]].

\bibitem{Aoki:2014dxa} 
  S.~Aoki, K.~Kikuchi and T.~Onogi,
  JHEP {\bf 1504}, 156 (2015)
  doi:10.1007/JHEP04(2015)156
  [arXiv:1412.8249 [hep-th]].

\bibitem{Suzuki:2015fka} 
  H.~Suzuki,
  PTEP {\bf 2015}, no. 4, 043B04 (2015)
  doi:10.1093/ptep/ptv036
  [arXiv:1501.04371 [hep-lat]].

\bibitem{Monahan:2015lha} 
  C.~Monahan and K.~Orginos,
  Phys.\ Rev.\ D {\bf 91}, no. 7, 074513 (2015)
  doi:10.1103/PhysRevD.91.074513
  [arXiv:1501.05348 [hep-lat]].

\bibitem{Endo:2015iea} 
  T.~Endo, K.~Hieda, D.~Miura and H.~Suzuki,
  PTEP {\bf 2015}, no. 5, 053B03 (2015)
  doi:10.1093/ptep/ptv058
  [arXiv:1502.01809 [hep-lat]].

\bibitem{Suzuki:2015bqa} 
  H.~Suzuki,
  PTEP {\bf 2015}, no. 10, 103B03 (2015)
  doi:10.1093/ptep/ptv139
  [arXiv:1507.02360 [hep-lat]].

\bibitem{Ramos:2015baa} 
  A.~Ramos and S.~Sint,
  Eur.\ Phys.\ J.\ C {\bf 76}, no. 1, 15 (2016)
  doi:10.1140/epjc/s10052-015-3831-9
  [arXiv:1508.05552 [hep-lat]].

\bibitem{Capponi:2015ahp} 
  F.~Capponi, L.~Del Debbio, A.~Patella and A.~Rago,
  PoS LATTICE {\bf 2015}, 302 (2016)
  [arXiv:1512.04374 [hep-lat]].

\bibitem{Datta:2015bzm} 
  S.~Datta, S.~Gupta and A.~Lytle,
  Phys.\ Rev.\ D {\bf 94}, no. 9, 094502 (2016)
  doi:10.1103/PhysRevD.94.094502
  [arXiv:1512.04892 [hep-lat]].

\bibitem{Christensen:2016wdo} 
  C.~Christensen and M.~Laine,
  Phys.\ Lett.\ B {\bf 755}, 316 (2016)
  doi:10.1016/j.physletb.2016.02.020
  [arXiv:1601.01573 [hep-lat]].

\bibitem{Fujikawa:2016qis} 
  K.~Fujikawa,
  JHEP {\bf 1603}, 021 (2016)
  doi:10.1007/JHEP03(2016)021
  [arXiv:1601.01578 [hep-lat]].

\bibitem{Hieda:2016lly} 
  K.~Hieda and H.~Suzuki,
  Mod.\ Phys.\ Lett.\ A {\bf 31}, no. 38, 1650214 (2016)
  doi:10.1142/S021773231650214X
  [arXiv:1606.04193 [hep-lat]].

\bibitem{Kamata:2016any} 
  N.~Kamata and S.~Sasaki,
  Phys.\ Rev.\ D {\bf 95}, no. 5, 054501 (2017)
  doi:10.1103/PhysRevD.95.054501
  [arXiv:1609.07115 [hep-lat]].

\bibitem{Taniguchi:2016tjc} 
  Y.~Taniguchi, K.~Kanaya, H.~Suzuki and T.~Umeda,
  Phys.\ Rev.\ D {\bf 95}, no. 5, 054502 (2017)
  doi:10.1103/PhysRevD.95.054502
  [arXiv:1611.02411 [hep-lat]].

\bibitem{Capponi:2016yjz} 
  F.~Capponi, L.~Del Debbio, S.~Ehret, R.~Pellegrini, A.~Portelli and A.~Rago,
  PoS LATTICE {\bf 2016}, 341 (2016)
  [arXiv:1612.07721 [hep-lat]].

\bibitem{Hieda:2017sqq} 
  K.~Hieda, A.~Kasai, H.~Makino and H.~Suzuki,
  PTEP {\bf 2017}, no. 6, 063B03 (2017)
  doi:10.1093/ptep/ptx073
  [arXiv:1703.04802 [hep-lat]].

\bibitem{Husung:2017qjz} 
  N.~Husung, M.~Koren, P.~Krah and R.~Sommer,
  EPJ Web Conf.\  {\bf 175}, 14024 (2018)
  doi:10.1051/epjconf/201817514024
  [arXiv:1711.01860 [hep-lat]].

\bibitem{Eller:2018yje} 
  A.~M.~Eller and G.~D.~Moore,
  arXiv:1802.04562 [hep-lat].

\end{thebibliography}
\end{document}